\documentclass[a4paper,usenatbib]{mnras}

\usepackage{newtxtext,newtxmath}

\usepackage[T1]{fontenc}
\usepackage{ae,aecompl}

\usepackage{graphicx}	% Including figure files
\usepackage{amsmath}	% Advanced maths commands
\usepackage{amssymb}	% Extra maths symbols
\usepackage{tabularx}
\usepackage{booktabs}
\usepackage{footmisc}
\usepackage{textcomp}
\usepackage{siunitx}
\usepackage{subfig}
\usepackage{caption}
\usepackage{diagbox}
\usepackage{slashbox}
\usepackage{grffile}
\usepackage{setspace}
\usepackage{hyperref}
\usepackage{lastpage}

\usepackage{booktabs} % Para linhas horizontais de alta qualidade
\usepackage{tabularx} % Para tabelas de largura fixa com colunas ajustáveis
\usepackage{siunitx} % Para formatar números e unidades
\usepackage{threeparttable} % Para adicionar notas de rodapé
\usepackage{nicematrix}
\usepackage{amsmath}
\usepackage{arydshln} % Pacote para linhas tracejadas
\newcolumntype{C}[1]{>{\centering\arraybackslash\fontsize{10}{12}\selectfont}m{#1}}

\title[Cl 0024+17 $\&$ MS 0451-03]{Boosting the evolutionary picture of Cl 0024+17 and MS 0451-03: A case study at intermediate-redshift}

\author[Costa et al.]{
A. P. Costa$^{1}$\thanks{E-mail: alissonpereira62@gmail.com},
A. L. B. Ribeiro$^{1}$, %\thanks{}, 
R. R. de Carvalho$^{2}$, and % and % \thanks{} and 
J. A. Benavides$^{3}$
\\
$^{1}$Laborat\'orio de Astrof\'{\i}sica Te\'orica e Observacional, Universidade Estadual de Santa Cruz-45650-000, Ilh\'eus-BA, Brazil\\
$^{2}$NAT - Universidade Cruzeiro do Sul / Universidade Cidade de S\~ao Paulo, 01506-000, Brazil \\
$^{3}$Department of Physics and Astronomy,
University of California (UCR), Riverside,
California, USA 
}

\date{Accepted 2024 October 18. Received 2024 September 20; in original form 2024 July 24}

\pubyear{2024}

\begin{document}
\label{firstpage}
\pagerange{\pageref{firstpage}--\pageref{lastpage}}
\maketitle

\begin{abstract}

In this work we improve the dynamic-evolutionary framework of two massive clusters at intermediate redshifts: Cl 0024+17 at $z \sim 0.4$ and MS 0451-03 at $z \sim  0.5$. The spectroscopic galaxy members were selected from Moran et al. (2007a), which combine optical and UV imaging with spectroscopy. Using a set of dynamic estimators with different approaches, our results show that both Cl 0024+17 and MS 0451-03 are non-relaxed systems with distinct dynamical configurations. Cl 0024+17 exhibits a disturbed kinematics, displaying significant gaps and a velocity dispersion profile suggesting a merger. This is confirmed by the presence of previously reported substructures and new ones identified in this study. MS 0451-03 appears less disturbed than Cl 0024+17, indicating by the significant segregation between late and early-type galaxies, with the latter occupying more central regions of the projected phase-space. However, five previously unobserved substructures and non-Gaussianity in the velocity distribution indicate that MS 0451-03 is also out of equilibrium. In both clusters, there are substructures infalling onto the systems, indicating key moments in their assembly histories and potential effects on the pre-processing of galaxies within these subgroups. This is suggested by the high percentage of early-type galaxies outside $R_{200}$ (approximately $83\%$) in the case of CL 0024+17. This work reinforces the importance of more detailed dynamical analysis of clusters to better characterize their evolutionary picture.
 
\end{abstract}


%%%%%%%%%%%%%%%%%%%%%%%%%%%%%%%%%%%%%%%%%%%%%%%%%%

%%%%%%%%%%%%%%%%% BODY OF PAPER %%%%%%%%%%%%%%%%%%

\section{Introduction}

In the hierarchical scenario of formation of larger structures ($\Lambda$CDM), galaxy clusters correspond to the collapse of the largest gravitationally bound over-densities in the initial density field and as such represent excellent laboratories for studying the physical processes operating in the universe at these scales (e.g. \citealt{1974ApJ...187..425P}; 
\citealt{2012ARA&A..50..353K}; \citealt{Migkas}).  It is important to note that
most of what we know about clusters comes from systems in the local universe (redshift $z\lesssim 0.1$, e.g. \citealt{2016AstBu..71..155T}), such as, for example, the environmental effects of these systems on the morphology and stellar content of galaxies, which makes them valuable sites to investigate how galaxies evolve in high-density regimes (\citealt{2006MNRAS.373..469B}; 
\citealt{2010ApJ...721..193P}; 
\citealt{2010A&A...520A..30M}; 
\citealt{2011MNRAS.416.2882M}; 
\citealt{2013ApJ...775..126H}; 
\citealt{2015MNRAS.448.1715J}; 
\citealt{Paranjape}; 
\citealt{Koulouridis}).

Clusters located at intermediate-to-high redshifts ($0.1 \lesssim $ z $\lesssim 1.0$ - although with recent observations this common range may be changed) are important pieces in the cosmic puzzle that can help us understand the previous stages of the structure formation process. As a striking example, clusters at intermediate-to-high redshifts exhibit a higher star formation rate than those at lower redshifts 
(e.g. \citealt{1978ApJ...219...18B}; \citealt{1984ApJ...285..426B}; \citealt{1538-3881-119-4-1562}). Despite the difficulty in identifying clusters, especially at the lower mass regime, the last two decades have witnessed the appearance of several reliable cluster catalogs. Using approximately 6,000 galaxies from CNOC2 (Canadian Network for Observational Cosmology field galaxy redshift survey) in the redshift range of 
$0.25-0.45$, \cite{2001ApJ...563..736C} conclude that groups with higher velocity dispersion largely act to suppress star formation, while groups with lower velocity dispersion are sites of strong merging and enhanced star formation which may lead to the formation of new massive galaxies at intermediate redshifts. The study by \cite{2001ApJ...563..736C} 
suggests a relation between star formation and clustering may be important for understanding the evolution of groups and field galaxies as a whole.

\cite{2008PASJ...60S.531G}, analysing an extinction-free indicator of galaxy star-formation activity, the mid-infrared $S_{15\mu m}/S_{9\mu m}$ flux ratio, in sixteen clusters in the redshift interval $0.9 < z <1.7$, observed that cluster member galaxies have low star-formation activity, indicating that star formation may have started at an even higher redshift. Another important piece of information, related to how galaxies evolve through time, comes from the strong-lensing dynamical analysis in seven galaxy groups observed with the Very Large Telescope (VLT) at intermediate redshifts ($0.3<z<0.7$), done by \cite{2013A&A...552A..80M}. They show that the virial mass of these systems are between $10^{13}\rm ~M_{\odot}$ and $10^{14}\rm ~M_{\odot}$, classifying them as groups or low mass clusters. Although two of these systems display signs of merging, the authors note a good agreement between the velocity dispersions estimated from the analysis of the group kinematics and the weak lensing measurements, concluding that the dynamics of baryonic matter are a good tracer of the total mass content in galaxy groups.
More recently, \cite{Feuillet} searching for alternative methods to classify galaxies at intermediate to high redshifts, aiming to overcome the limitations of the standard BPT diagram (which has a redshift limit of z $\lesssim 0.57$) studied a sample of galaxies coming from \textit{Sloan Digital Sky Survey} (\citealt{sdss}) Data Release 12. They derived new diagnostic diagrams that allow them to classify galaxies up to z $\lesssim 1.07$. Their methodology, using samples in intermediate redshfits, was able to provide properties of active galaxy nuclei (AGN), in addition to enabling a more complete view of the number of AGN in the universe, which is important for the study of galaxy evolution.

Although previous studies have contributed significantly to the validation of global physical trends (due to the statistical extent of samples that reduces possible biases, for example), they may ignore individual variations that can be fundamental to understanding the diversity involved in physical processes at work in these systems. From this perspective, case studies, as proposed in this investigation, are ideal for exploring the complexity and nuances of specific systems, in addition to providing a more detailed and contextualized analysis (e.g. ~\citealt{Cerda}; ~\citealt{Shao}).
Taking this into consideration, we perform a comprehensive dynamical analysis of two clusters at intermediate-redshift, namely: Cl 0024+17+17 ($z = 0.39$) and MS 0451-03 ($z=0.54$) from \cite{moran2007wide} and \cite{moran2007b}. Our purpose is to fully characterize the dynamic picture of these two systems. We chose to study these clusters because besides extending 
and consequently completing previous analysis by \cite{moran2007wide, moran2007b}, they have two important characteristics: (i) they exhibit significant differences in X-ray emission; and (ii) they have very different masses, MS 0451-03 being $\sim 50\%$ more massive than Cl 0024+17. As discussed by \cite{2019MNRAS.484.2807L},  physical properties of the X-ray gas are an important dynamical proxy for galaxy clusters, which can be segregated into cool-core (CC, relaxed systems with few signs of perturbation) and non-cool core (NCC).
Furthermore, clusters studied in the local universe have shown a strong correlation between the X-ray observables and the dispersion of velocities of their member galaxies (e.g.~\citealt{Ortiz2004} and ~\citealt{nastasi2014}). Also, environmental effects on galaxies may be related to the mass of the system where they are located \citep{Wetzel2012,2019MNRAS.483L.121N,Oman2021}. All this suggests that revisiting the study of these two clusters can contribute to the understanding of the
formation processes of clusters and how these affect the evolution of galaxies.

This paper has the following structure. In Sec. \ref{sec2} we present our
clusters and galaxies sample. In Sec. \ref{sec3}, we present the different methods
used to robustly classify the clusters’ dynamical picture, and finally, in Sec. \ref{sec4} we summarize and discuss our analysis results.
Throughout this paper, we adopt a $\Lambda$CDM cosmology with $H_{0}$, $\Omega_{M}$ and $\Omega_{\Lambda}$ from \cite{planck2018} results.

\section{DATA}\label{sec2}

\subsection{Overall}
\label{subs1}

Cl 0024+17 at $z=0.39$ and MS 0451-03 at $z=0.54$ are two massive clusters with virial mass of $\rm M_{200}=8.7\times10^{14}\rm ~M_{\odot}$ and $\rm M_{200}=1.4\times10^{15}\rm ~M_{\odot}$, respectively. Their sizes are $\rm R_{200} = 0.95 ~Mpc$ and $\rm R_{200} = 1.45 ~Mpc$, respectively\footnote{Typically Virial quantities, such as mass or radius ($M_{200}$ and $R_{200}$) are determined
by the radius where the average density equals 200 times the critical density of the Universe ($\rho_c = 3H^2 /8\pi G$).}, 
which were determined following \cite{1997ApJ...478..462C}. Member galaxies around Cl 0024+17 and MS 0451-03 are retrieved from \cite{moran2007wide} who combine optical and UV imaging from the Hubble Space Telescope (HST) and The Galaxy Evolution Explorer (GALEX). They also provide spectroscopy and magnitudes (bands I and V only) for all member galaxies, in addition to approximately 83$\%$ of the [OII]3727~\AA{} equivalent width values. These two systems are part of a study to investigate the global properties of early-type passive galaxies in clusters, as well as being a contribution to a long-term project to study the evolution of galaxies in fields considered extensive ($\sim 10$~Mpc) (\citealt{moran2007b} and \citealt{moran2007wide}).

While MS 0451-03 represents one of the highest X-ray luminosity intermediate-redshift cluster (\citealt{2003ApJ...598..190D}), Cl 0024+17 does not show a significant X-ray component when observed with XMM (X-ray Multi-Mirror Mission) (\citealt{2005A&A...429...85Z}), and has been the subject of many studies since its discovery by 
\cite{Humason&Sandage}, for example, \cite{1998ApJ...498L.107T}, \cite{2000ApJ...534L..15B}, \cite{2006ApJ...642...39C} and \cite{2009NewA...14..666R}. The fact that MS 0451-03 has an X-ray luminosity seven times higher than Cl 0024+17 ($5.3\times10^{11}\,\,L_{\odot}$ and $7.6\times10^{10}\,\,L_{\odot}$, respectively) suggests a significant difference 
in the Intra-Cluster Medium (ICM) density, maybe implying a different evolution of infalling galaxies in MS 0451-03 when compared to Cl 0024+17. The selection of member galaxies in Cl 0024+17 means galaxies in the two prominent peaks of the cluster redshift distribution, following \cite{2002A&A...386...31C}.  This encompasses a redshift range from $z = 0.373$ to $z = 0.402$, which corresponds to a line-of-sight (\textit{los}) velocity range of $[-3857.12; 2309.98]$ km/s and overall velocity dispersion ($\sigma_{cl}$) of $871.50$ km/s, totaling 508 galaxies. The ratio between the values $|\sigma_{cl}/V_{mean}|\sim 3.8$.
For MS 0451-03, all galaxies in the redshift interval $0.52<z<0.56$ (with equivalent \textit{los} velocity range of $[-4695.84; 2353.29]$ km/s) were considered members of the system, resulting in 319 galaxies. Furthermore, the $|\sigma_{cl}/V_{mean}|$ ratio for MS 0451-03 (with $\sigma_{cl}=920.01$ km/s), also interpreted as a measure of velocity variability, indicates a value of $\sim 3.0$. The redshift distributions for both clusters and more details on general member selection are found in \cite{moran2007wide}.

Morphological classification is available for all galaxies brighter than absolute magnitude in the $V$-band, $M_{V}=-19.5$ (limiting magnitude for a galaxy to be considered a member, \citealt{moran2007wide}). This corresponds to $I=22.1$ for MS 0451-03 and $I=21.2$ in Cl 0024+17. Thus, morphological type was assigned according to the scheme defined in the Medium 
Deep Survey introduced by \citealt{1996ApJ...471..694A}: T=-2=star, -1=compact objects, 0=E, 1=E/S0, 2=S0, 3=Sa+b, 4=S, 5=Sc+d, 6=Irr, 7=Unclassified, 8=Merger objects, 9=Fault. All galaxies assigned with the T=0, 1, 2 are identified as early-types or E+S0s and all galaxies with the assignments T=3, 4, 5 are identified as spirals in agreement with 
\cite{moran2007wide}. For Cl 0024+17 the corresponding $\rm M_{V}^{*} \,(\sim -21.27)$ is approximately $I\sim19.5$ (\citealt{1997ApJS..110..213S}). In general, both clusters have morphological information with a more rigorous distinction up to $\rm M_{V}^{*}+1.6$ and a more general and broad classification extending up to $M_{V}^{*}+3.0$ (\citealt{moran2007wide}).

\subsection{Categorizing galaxy members according to luminosity regime }

The evolution of the galaxy population in clusters is still an ongoing problem. One essential characteristic of a galaxy population is its luminosity function, whose evolution tell us how this population evolve as a result of star formation, stellar aging and galaxy merging. However, the study of the cluster luminosity function at intermediate to high redshifts has been limited to the galaxies in red sequence, usually used to detect and define the cluster (e.g. \citealt{2009MNRAS.399.1858L}, \citealt{2012ApJ...745..106L}). These small datasets have caused disagreements in interpreting the behavior of the luminosity function of those systems. For example, while some authors find that the faint end of the high redshift clusters luminosity function is decreasing (e.g. \citealt{2009ApJ...700.1559R}), others still see a flat faint end (e.g. \citealt{2013MNRAS.434.3469D}). 

\begin{figure} %10.3cm, height=10cm
  %\hspace*{-0.4in}
  \centering
        \includegraphics[height=8cm, width=8cm]{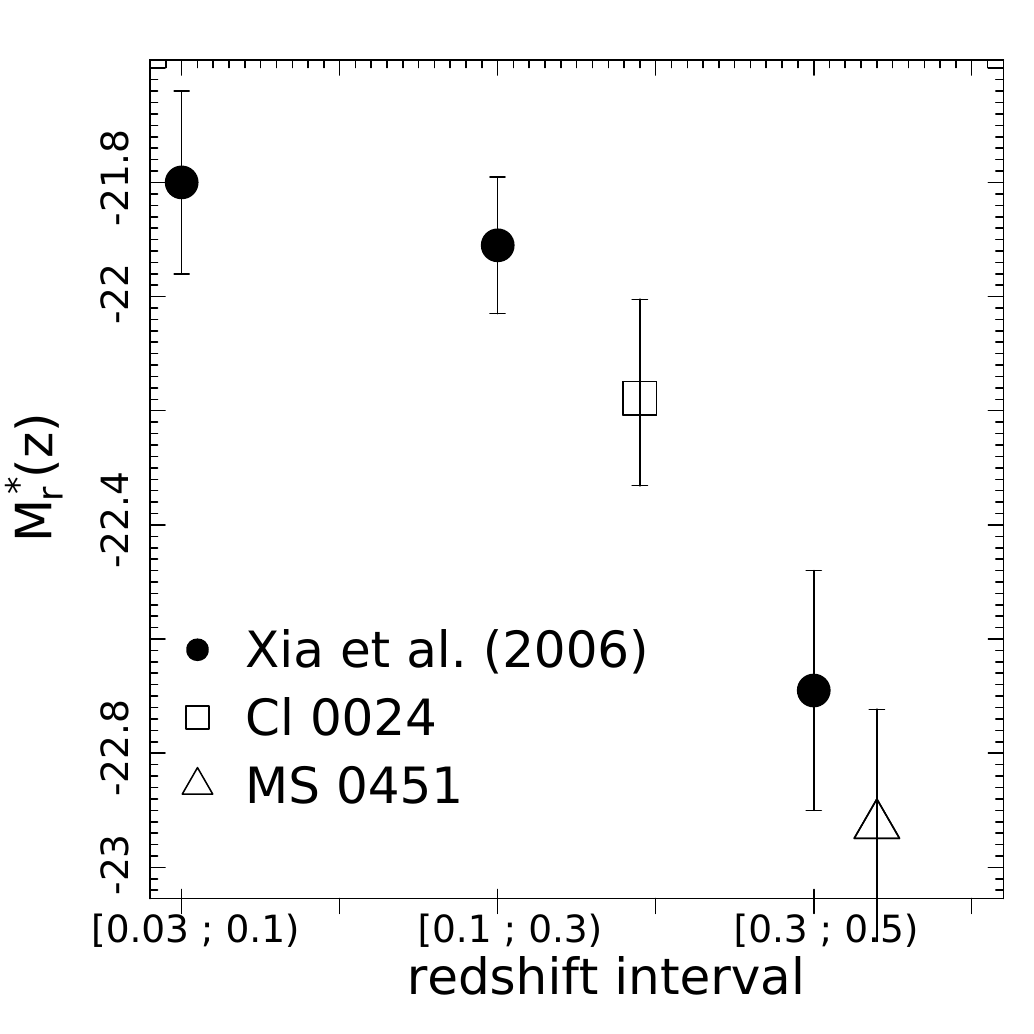} \quad 
       \caption{Evolution of the characteristic magnitudes of Cl 0024 and MS 0451 in the r band estimated from data provided by \protect\cite{2006ApJ...652..249X}. 
       The values of the characteristic magnitudes are $\rm M^{*}_{r}= -22.18 \pm 0.16$ for Cl 0024 and 
       $\rm M^{*}_{r}= -22.91 \pm 0.18$ for MS 0451.}

     \label{fig_lf}
\end{figure}

Considering these issues, together with the lack of information on absolute magnitude completeness in V-band of both clusters, we are not fitting the luminosity function to the available data, since the results can be misleading. Thus, to explore the dynamics of the galaxies segregated in luminosity (in future sections), that is, in the \textit{Bright} (+L) and \textit{Faint} (-L) regimes, we adopt the following strategy. Initially we verify the evolution of the characteristic magnitudes, $\rm M^{*}$, of Cl 0024+17 and MS 0451-03 in r band. To do so, we use data from the studies of clusters at intermediate to high-redshift coming from \cite{2006ApJ...652..249X}. The result is shown in Figure \ref{fig_lf}. 
% \cite{2007ApJ...659.1106M} and \citealt{1999AJ....118..719D}

Using a cubic spline, we estimate the characteristic absolute magnitude $\rm M^{*}$ for the two clusters in the r band and assuming the tendency between $\rm M^{*}$ and z. With this procedure, we are assuming that both clusters follow an average trend observed in larger samples, which can occasionally result in possible biases for $\rm M^{*}$. In light of this, we find the following results: $\rm M^{*}_{r}= -22.18 \pm 0.16$ for Cl 0024+17; and $\rm M^{*}_{r}= -22.91 \pm 0.18$ for MS 0451-03. These values show that the difference between the characteristic magnitudes between Cl 0024+17 and MS 0451-03 is $\sim 0.7$ mag. 
Taking into account that \cite{2003ApJ...591...53T} reported that at redshifts around $\sim0.4$, changes on the order of $\sim0.5$ mags do not substantially alter the results achieved in regards to luminosity,
we assume that this difference in $\rm M^{*}_{r}$ ($\sim 0.7$ mag) stays the same in V-band. Also, the characteristic absolute magnitude in V band, $\rm M^{*}_{V}$, corresponds to $\sim$ 19.5 in I for Cl 0024 (what matches $\rm M^{*}_{V} \sim -21.27$, \citealt{1997ApJS..110..213S}). The $\rm M^{*}_{V}$ value for MS 0451-03 will be given by adding 0.7 mag to the $\rm M^{*}_{V}$ value of Cl 0024+17. Therefore, for Cl 0024+17, $\rm M^{*}_{V} \sim -21.27$, while for MS 0451-03, $\rm M^{*}_{V} \sim -21.97$.

Finally, we consider galaxies belonging to the +L regime those that have absolute magnitude $\rm M_{V} \leq M^{*}_{V}+1$. To probe the -L regime, in order to cover the same portion of the luminosity function and maintain coherence in the comparison of the galaxy populations of both systems, we must find a number $k$ that satisfies $\rm M_{lim,\,V} = M_{V}^{*} + k$ for Cl 0024+17 and MS 0451-03. For this purpose, we use a limiting magnitude of $\rm M_{lim,\,V} =-19.5$ (see subsection \ref{subs1} for more details on the value considered for $\rm M_{lim,\,V}$). That done, the values of $k$ satisfying the previous equality for Cl 0024+17 and MS 0451-03 are $k_{Cl} = 1.77$ and $k_{MS}=2.47$, respectively. Therefore, the -L regime is defined as $\rm M^{*}_{V} + 1 < M_{V} \leq M^{*}_{V} + k$ and from now on when we refer to all galaxies, that means galaxies brighter than $\rm M_{lim,\,V} =-19.5$, for both clusters.

At this point we emphasize that, although our strategy is the most appropriate for this situation, the results obtained in the following sections for the +L and -L samples must be understood with reasonable caution, since due to the scarcity of information on characteristic magnitudes of clusters at similar redshifts compared to Cl 0024+17 and MS 0451-03, it is unfeasible to consider more rigorous criteria for determining $M^{*}$.

In view of the evidence that morphology plays an important role in the characterization of galaxy clusters (e.g. \citealt{beisbart2001morphological}; \citealt{Gnedin_2003}; \citealt{fasano2015morphological}), we split the population of galaxies of both clusters in two groups. One encompasses early (E, E/S0, S0) galaxies and the other corresponds to the late type galaxies represented by S, Sa+b and Sc+d.
These subsamples (in luminosity and morphology) will be studied in Subsections \ref{subvdp} and \ref{subpps}.

\section{Results and Analysis} % Dynamical Study on Cl 0024 and MS 0451
\label{sec3}

The purpose of this section is to constrain the dynamical stage of the clusters Cl 0024+17 and MS 0451-03 in a systematic fashion. For this we introduce a set of tools to characterize the dynamical stage of a galactic system which should be understood in the most unified way possible, since in isolation they reflect only a single aspect of the complexity of a cluster of galaxies and may mislead future conclusions. 
The aspects and perspectives analyzed are:

\begin{itemize}

\item Velocity Distribution of the member galaxies. Although this is a line-of-sight projection, we may have some indication of the dynamical stage of the system. We assume that a gaussian distribution indicates how far from relaxation the system is (e.g. \citealt{1977ApJ...214..347Y}; \citealt{1993AJ....105.1596B}; \citealt{2013MNRAS.434..784R}); \\

\item Velocity dispersion profiles (VDP). The expectation is that the VDP shape may reveal different dynamical mechanisms in progress (e.g. \citealt{1996ApJ...472...46M}; \citealt{biviano2004eso}; \citealt{2012MNRAS.421.3594H};  \citealt{2018MNRAS.473L..31C}); \\

\item Projected Phase-Space (PPS). In theory, the PPS is a space of all possible states of a dynamical system. From this viewpoint it may indicate segregation and/or evolutionary effects (e.g. \citealt{2013ApJ...768..118N}); \\

\item Spatial Distribution Analysis. The presence of substructures in the projected galaxy distribution may be indicative of an active history of past or ongoing mergers of smaller groups and clusters that affect the evolution of galaxies by enhance and/or quench their star formation (e.g. \citealt{2006PASP..118..517B}; \citealt{2014ApJ...783..136C}; \citealt{2018A&A...610A..82E}; \citealt{2019arXiv190109198F}).

\end{itemize}

\subsection{Velocity distribution}
\label{veldisp}

It is well known that in general the evolution of galaxy clusters passes through two major stages: one called violent relaxation (\citealt{1967MNRAS.136..101L}) where the evolution of the system is controlled by a collective potential; and through two-body encounters, which leads to the kinetic energy equipartition  and mass segregation. 
One of the key outcomes of these processes is that the 3-D velocity distribution of galaxies can be modeled by a Maxwell-Boltzmann distribution. However, since we only access the velocity component projected along the line of sight, \textit{N}-body simulations and phenomenological studies (e.g. \citealt{1990AJ....100...32B}; \citealt{2003ApJ...595...43M}; \citealt{2005NewA...10..379H}; \citealt{2009ApJ...702.1199H}; \citealt{2011MNRAS.413L..81R}; \citealt{2012A&A...540A.123E} and references therein) support that such velocity component is best described by a Gaussian distribution, when the system is in dynamic equilibrium.

We investigate the velocity distribution of Cl 0024+17 and MS 0451-03 using two important indicators whose results can be seen in a complementary way. The \texttt{Mclust} analysis (\citealt{fraley2012mclust}) (a package in \texttt{R}\footnote{An open-source free statistical environment developed under the GNU GPL \cite{ihaka1996r}.}) searches for an optimal number of gaussian modes in the distribution considering a finite distribution mixture, assigning to each galaxy a probability of belonging to a specific mode (e.g. \citealt{2011MNRAS.413L..81R}; \citealt{2012A&A...540A.123E} and \citealt{10.1093/mnras/staa2464}). The method is based on a search of an optimal model for the clustering of the data among models with varying shape, orientation and volume. It finds the optimal
number of components and the corresponding classification (e.g. ~\citealt{Carvalho_2017}).

We also use Anderson-Darling (AD) to measure gaussianity in velocities distributions. The AD test (\citealt{stephens1974edf}) or $A^{2}$ statistic has been systematically used in recent years to probe the velocity distribution of cluster galaxies, for example, \cite{Hou_2009}, \cite{2012MNRAS.421.3594H}, \cite{2018MNRAS.475.4704R} and \cite{2023A&A...676A.127D}. The AD test gives more weight to the tails of the distribution and is considered a rather sensitive method
to deviations from gaussianity (e.g. \citealt{Hou_2009}). For both tests we adopted a confidence level of $95\%$ (e.g.~\citealt{2023A&A...676A.127D}).

If we take all galaxies, regardless of their location, morphology and luminosity, according to \texttt{Mclust} and AD measurements, Cl 0024+17 is a system with Non-Gaussian (NG) velocity distribution with $99.8\%$ reliability\footnote{Reliability is calculated using one thousand realizations (bootstrap) of the velocity distribution -- see \cite{1538-3881-154-3-96}.}, suggesting the presence of two modes, and $p$-value $< 3.10^{-12}$ for AD test. For MS 0451-03 the velocity distribution is considered Gaussian (G) according to \texttt{Mclust} with 99$\%$ reliability and also G by AD test with a $p$-value$=0.02$.

Considering the extension over which the two clusters are probed (see section \ref{subs1}), we also evaluate the velocity distribution of two distinct regions in terms of $R_{200}$. Our intention is to evaluate the velocity distribution of the systems in more detail and check whether there is any region that exerts a significant influence on the entire respective velocity fields. The results are in Tables \ref{tabela1} and \ref{tabela2}.

\begin{table}
\begin{center}
\caption{Results for Cl 0024+17 of the \texttt{Mclust} and AD tests. The upper part of the table refers to results obtained using the galaxies in the inner region of the cluster ($\rm R \leq R_{200}$) up to the limiting magnitude, $\rm M_{lim,\,V} =-19.5$. The lower part displays results for $\rm R > R_{200}$.}
\begin{tabular}{lcc}
\hline
$\rm R \leq R_{200}$ (inner) & Test & Results (\texttt{Mclust}, $p$-value)\\
\midrule
all galaxies & \texttt{Mclust} & NG (100$\%$)\\
all galaxies & AD & NG (p-$value < 10^{-6}$) \\
\\
\hline
$\rm R > R_{200}$ (outer) & Test & Results (\texttt{Mclust}, $p$-value)\\
\midrule
all galaxies & \texttt{Mclust} & NG (100$\%$)\\
all galaxies & AD & NG (p-$value < 10^{-5}$)\\
\\
\hline
\end{tabular}
\label{tabela1}
\end{center}
\end{table}

%%%%%%%%%%%%%%%%%%%%%%%%%%%%%%%%%%%%%%%%%%%%%%%%%%%%%%%%%%%%%%%%%%%%%%%%%%%%%%%%%%%%%%%%%%%%%%%%%%%%%%%%%%%%%%%%%%%%%%%%%%%%%%%%%%%%
%%%%%%%%%%%%%%%%%%%%%%%%%%%%%%%%%%%%%%%%%%%%%%%%%%%%%%%%%%%%%%%%%%%%%%%%%%%%%%%%%%%%%%%%%%%%%%%%%%%%%%%%%%%%%%%%%%%%%%%%%%%%%%%%%%%%

\begin{table}
\begin{center}
\caption{Results of the \texttt{Mclust} and AD tests for MS 0451-03. Regions and limiting magnitude are similar to that done in Cl 0024+17.}
\begin{tabular}{lcc}
\hline
$\rm R \leq R_{200}$ (inner) & Test & Results (\texttt{Mclust}, $p$-value)\\
\midrule
all galaxies & \texttt{Mclust} & G (100$\%$)\\
all galaxies & AD & G (p-$value = 0.33$) \\
\\
\hline
$\rm R > R_{200}$ (outer) & Test & Results (\texttt{Mclust}, $p$-value)\\
\midrule
all galaxies & \texttt{Mclust} & NG (100$\%$)\\
all galaxies & AD & NG (p-$value < 10^{-3}$)\\
\\
\hline
\end{tabular}
\label{tabela2}
\end{center}
\end{table}

From Table \ref{tabela1} (Cl 0024+17), we notice that, using all galaxies, \texttt{Mclust} and AD indicates that inner and outer regions have NG velocity distributions. This is in agreement with \cite{moran2007wide}, which (using another approach and with less rigor regarding the limiting magnitude of galaxies) finds that at least two substructures are infalling into the $R_{200}$ region. 
From this point of view, we can safely argue that Cl 0024+17 is a dynamically disturbed system. 
In MS 0451-03 (Table \ref{tabela2}) the two tests indicate that the central region behaves like a relaxed system, while the outer region experiences possible physical processes that make it non-Gaussian (disturbed).

\begin{figure} %10.3cm, height=10cm
  %\hspace*{-0.4in}
	\centering
        \includegraphics[height=7cm, width=7cm]{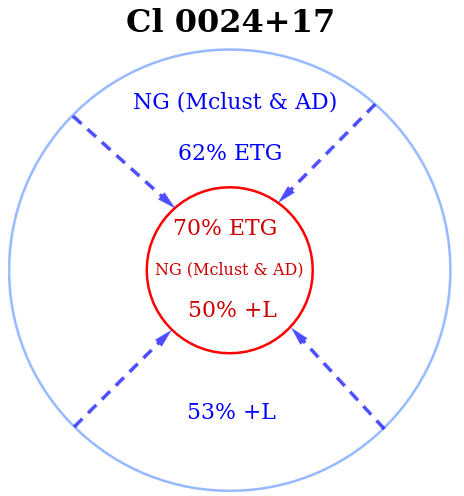} \quad
        \hspace*{-0.08in}
	\includegraphics[height=7.2cm, width=7cm]{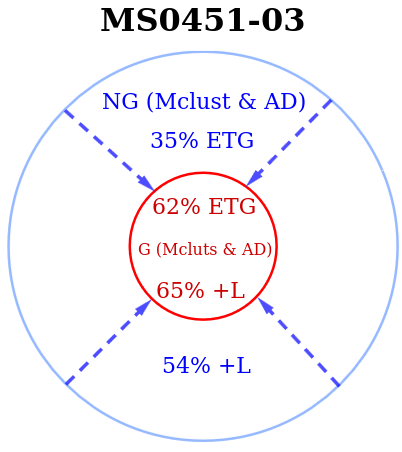} %\quad
       \caption{Schematic diagram of the fractions of objects presents in Cl 0024+17 and MS 0451-03 divided in two main regions by respective $R_{200}$.}
     \label{figesche}
\end{figure}

In Figure \ref{figesche}, we show a schematic representation for both clusters illustrating where the tests were applied. In each diagram we also show the fraction of galaxies $+L$ and early (and consequently $-L$ and late) in each of the regions. The intention is to verify the existence of unusual behavior related to the fraction of some populations of galaxies, since greater fraction of most luminous and ETGs in the central regions are important indicator not only of galaxy evolution, but also provides clues to the dynamical state of clusters.

As seen in Figure \ref{figesche}, in the inner region of Cl 0024+17 (red circle), ETGs and $+L$ systems represent the dominant population of galaxies, being distributed more centrally. This predominance is also true in the external region. Such fractions in the outskirts may, possibly, be related to the infall of smaller groups from filaments or general field around the cluster, inasmuch as  in subclumps outside the cluster core are the places where the transition from blue to red galaxies occurs most frequently (\citealt{Kodama}).

The diagram for MS 0451-03 reveals that although the fraction of $+L$ systems is approximately similar when comparing inner and outer regions, the fraction of ETGs differs significantly (a 27\% difference). These results may suggest a more evolved cluster (more morphological segregation towards ETGs), but caution must be exercised with respect this statement and then we will analyze another perspective of the velocity distribution to complement this part of the investigation.

To do this, we investigate the presence of significant gaps in the \textit{los} velocity distribution (e.g. \citealt{2012A&A...540A.123E}). We perform the gap analysis, similar to the one employed in \cite{2006A&A...449..461B}, \cite{2006A&A...455...45G} and \cite{doi:10.1093/mnras/stw1114}, to estimate the kinematic complexity of the velocity field around each cluster, based on the assumption that systems with more gaps may be at a more complex dynamical stage. To begin the analysis we first order the \textit{los} velocities so that the i-th gap in the velocity distribution is given by $g_{i}=v_{i+1} + v_{i}$. The weighting of a given gap is determined by $w_{i}=i(N-i)$. Thus, the value of the weighted gap is given by $\sqrt{w_{i}\,g_{i}}$. The weighted gaps are normalized by the central mean value of the weighted gaps distribution, given by

\begin{equation}
MM=\frac{2}{N}\sum_{i=N/4}^{3N/4}\,\sqrt{w_{i}\,g_{i}}\, \,\,.
\end{equation}\\

\begin{figure} %10.3cm, height=10cm
   \centering
   \vspace*{-0.2in}
        \includegraphics[height=8.4cm, width=8cm]{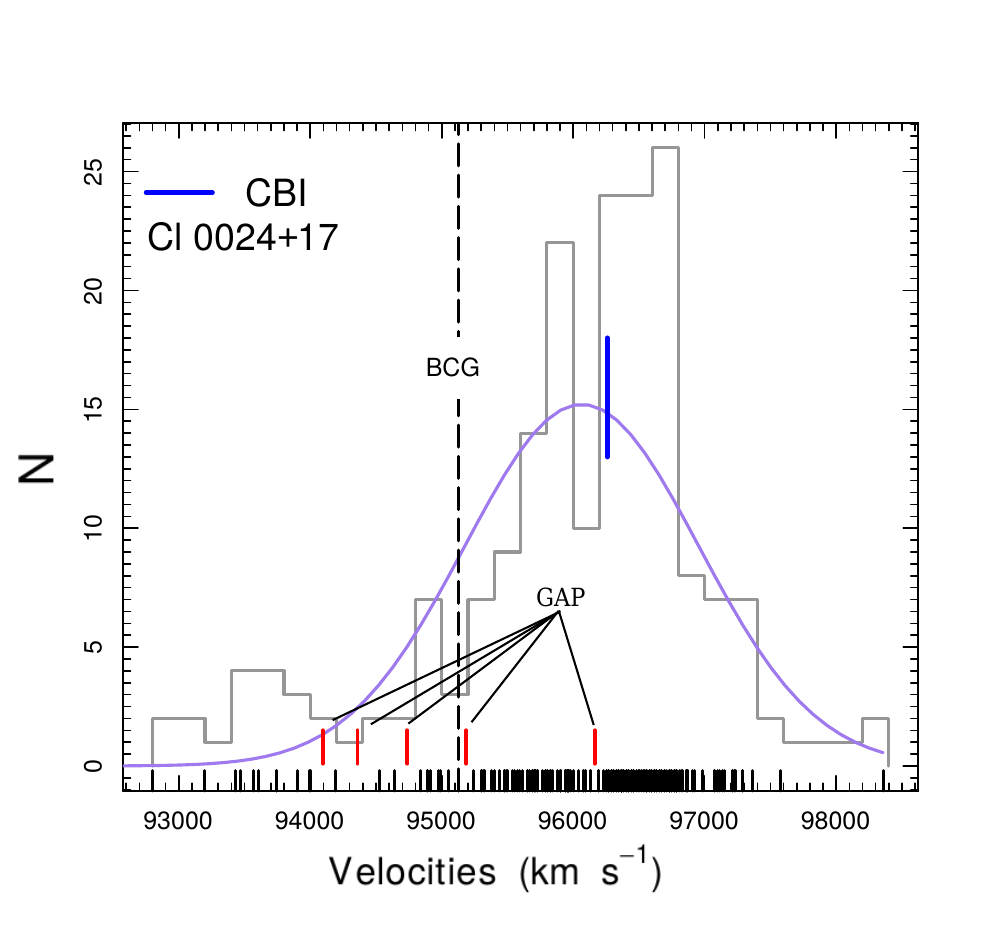}\quad 
    \vspace{0in}
	 \includegraphics[height=8.4cm, width=8cm]{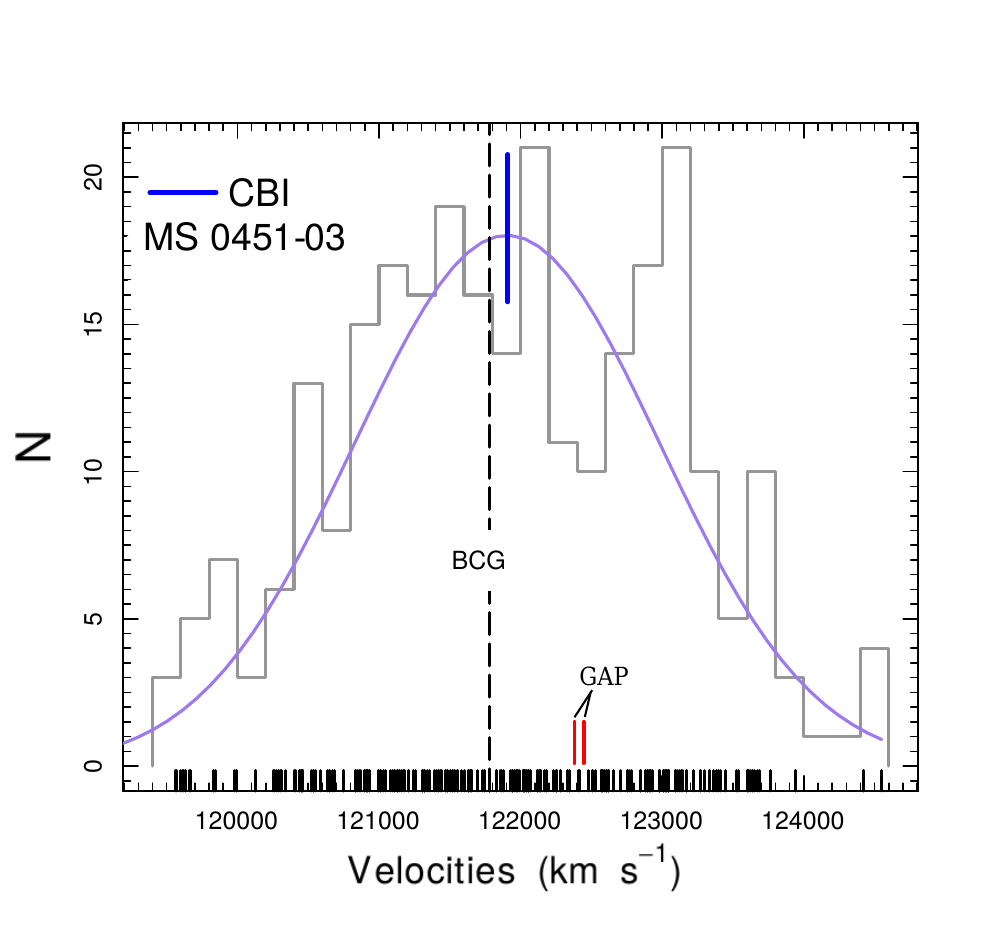} %\quad
       \caption{Histograms of the radial velocity distributions of clusters Cl 0024+17 (top) and MS 0451-03 (bottom), along with the superposition of the normal distribution with the same mean and standard deviation of the data. The black dashed 
       line represents the position of BCG in both fields. The dashes in red represent the position of the gaps found and the blue line the position of the estimate given by the $C_{BI}$.
       At the bottom of each graph we show a rug plot where we see the position of the galaxies in each distribution.}   
     \label{fig06}
\end{figure}

\noindent Here, we look for normalized gaps that are greater than $3.0$, since gaps larger than this appear in only $\sim 0.2\%$ of the cases (e.g. \citealt{wainer1978gapping} and \citealt{beers1991dynamical}). Thus, by choosing as significant gaps values greater than $3.0$, we ensure that the detected gaps are highly likely to be real and not due to fluctuations in the distributions. The results of this analysis are shown in Figure \ref{fig06}. In addition to the gaps we find, we also estimate the mean of the distribution using the robust estimator $C_{BI}$, discussed in \cite{1990AJ....100...32B}, along with the position of the brightest cluster galaxy (BCG).

At the top of Figure \ref{fig06}, the one corresponding to Cl 0024+17, we see a considerable number of gaps in the \textit{los} velocity distribution (indicating a complex velocity field), and also a large distance between the position of the BCG and the location estimator $C_{BI}$ for this cluster ($\Delta V \sim\, 1000$ km/s). In MS 0451-03, bottom of Figure \ref{fig06}, the distribution exhibits two gaps that are so close to each other that can be considered as one. This single gap indicates an important difference in the velocity field with respect to Cl 0024+17. Regarding the position of the BCG and the location estimator $C_{BI}$ we see that the difference between the two is very small ($\Delta V \sim\, 65$ km/s), suggesting a less disturbed dynamical configuration for MS 0451-03.

\subsection{Velocity dispersion profiles - VDPs}
\label{subvdp}

\begin{figure}
\hspace*{-0.1in}
%\begin{center}
	\includegraphics[height=6cm, width=8cm]{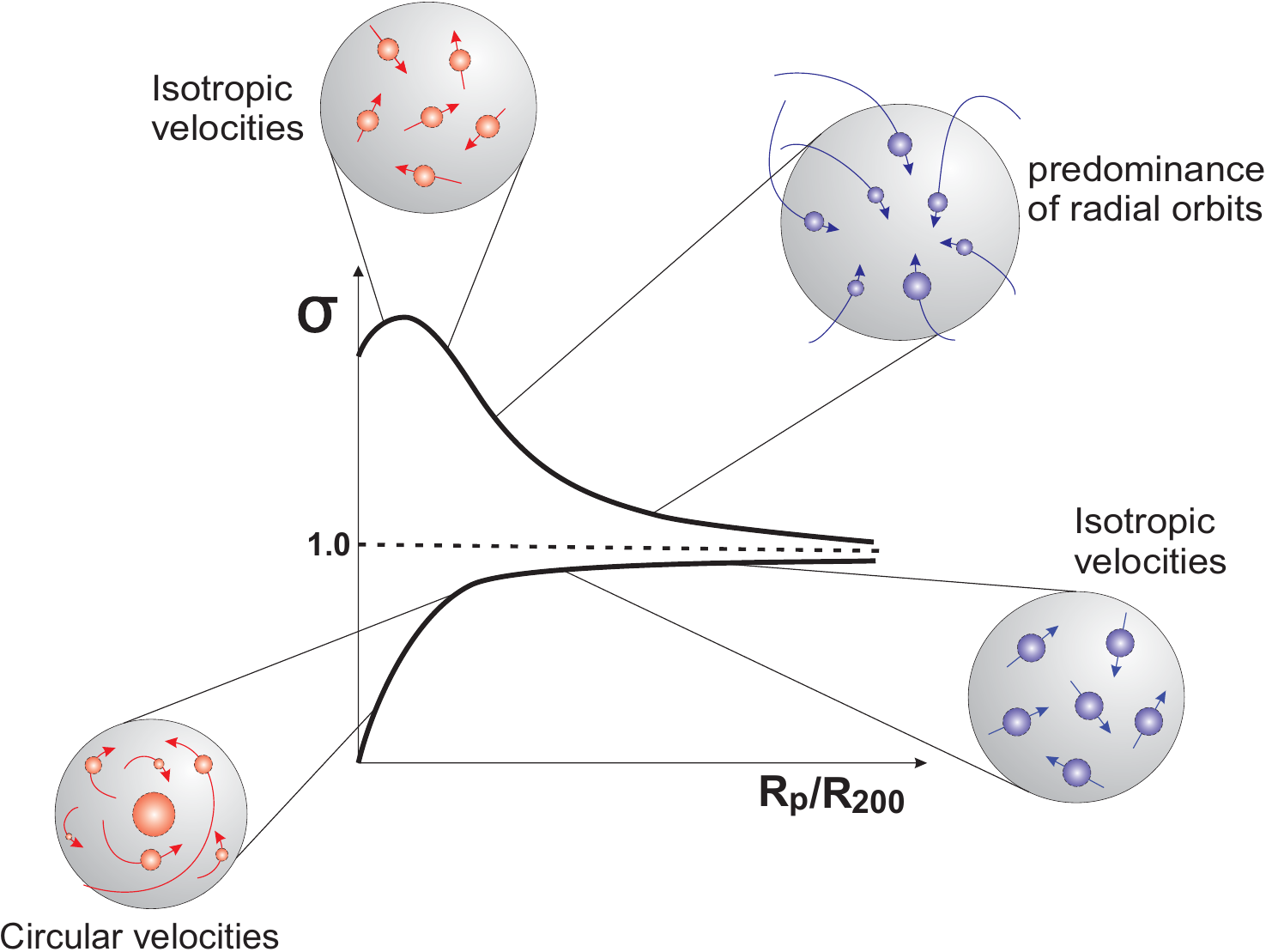}
\caption{A schematic diagram summarising the types of galaxy velocity predominance according to the characteristics of the VDPs as suggested by 
\protect\cite{1998ApJ...505...74G}. The scheme illustrates the most common cases of VDPs, that is, rising from the center (or decreasing to the center) followed by flat trend, decreasing from the center (peaked or not) with flat behavior next, or a VDP totally flat (dashed line).} 
\label{fig_schema}
%\end{center}
\end{figure}

Theoretical estimation of the velocity dispersion profile (VDP) requires breaking down the degeneracy between the anisotropy parameter and the system mass profile in the Jeans equation (see e.g. \citealt{1982MNRAS.200..361B} and \citealt{1987ApJ...313..121M}).  However, determining the VDP using robust statistical methods proves to be a particularly useful alternative to solve this singular problem, since theoretical and phenomenological studies (e.g. \citealt{1998ApJ...505...74G}; \citealt{2003MNRAS.343..401L}; \citealt{2004A&A...424..779B}; \citealt{2005MNRAS.362...95M}; \citealt{2009MNRAS.399..812W}; \citealt{2010MNRAS.401.2433M}; \citealt{2017A&A...606A.108C}; \citealt{2018arXiv181010474T} and references therein) show a very good agreement between both approaches. In Figure \ref{fig_schema} we represent, through \cite{1998ApJ...505...74G} studies, the main behaviors of VDPs with the physical interpretations associated with such characteristics. It is important to say that such characteristics are not exclusive to the VDP shapes in which they are represented in Figure \ref{fig_schema} and can be observed in other shapes of VDPs not studied by \cite{1998ApJ...505...74G}. These results are widely used in the study of VDPs (e.g. \citealt{2013MNRAS.436.2639F}; \citealt{2015MNRAS.453.2718W}; \citealt{2015A&A...579A...4G}; \citealt{2019MNRAS.483.4354S}; \citealt{2019MNRAS.483L.121N}; \citealt{2019arXiv190802277G}).

From this perspective, we obtain the VDPs to Cl 0024+17 and MS 0451-03 using a robust scale biweight estimator described in \cite{1990AJ....100...32B}
combined with the bias correction derived by \cite{Ferragamo} for small samples.
After a bit of algebra, we obtain (Eq. \ref{new_vdp}) a new expression for calculating VDPs in a robust way for samples with a minimum of eight galaxies.

%%%%%%%%%%%%%%%%%%%%%%%%%%%%%%%%%%%%%%%%%%%%%%%%%%%%%%
%%%%%%%%%%%%%%%%%%%%%%%%%%%%%%%%%%%%%%%%%%%%%%%%%%%%%%

\begin{eqnarray}
\text{\Large{$\sigma$}}_{BI} & = & \frac{\left[(1+B+D^{\beta})^{2} \sum\limits_{|u_{i}|<1} (x_{i}-M)^{2} (1-u_{i}^{2})^{4}\right]^{1/2}}{n^{-1/2}\left|\sum\limits_{|u_{i}|<1}(1-u_{i}^{2})(1-5u_{i}^{2})\right|\times (N_{gal}-1)^{\beta} }
\label{new_vdp}
\end{eqnarray}\\

%%%%%%%%%%%%%%%%%%%%%%%%%%%%%%%%%%%%%%%%%%%%%%%%%%%%%%%%%%%%%%%%%%%%%%%%%%%%%%%%%%%%%%%%%%%%%%%%%%%%%%%%%%%%%%%%%%%%%%%

In Eq. (\ref{new_vdp}), $B$, $D$ and $\beta$ are the parameters coming from \cite{Ferragamo} bias correction, while $x_{i}$, $u_{i}$ and $M$ are variables coming from \cite{1990AJ....100...32B}. 

To construct the VDPs, we opted for the methodology described by \cite{2018MNRAS.473L..31C}, instead of using techniques that perform parametric smoothing in velocity dispersion values. We argue that this type of tool can present challenges that could make the results obtained by interpreting VDPs misleading. For example, parametric smoothing can lead to loss of important information about the data used (if the smoothing parameter is excessive); distortion of significant peaks, changing the actual shape of the distribution; or even sampling bias in small data sets.

%%%%%%%%%%%%%%%%%%%%%%%%%%%%%%%%%%%%%%%%%%%%%%%%%%%%%%
%%%%%%%%%%%%%%%%%%%%%%%%%%%%%%%%%%%%%%%%%%%%%%%%%%%%%%

%%%%%%%%%%%%%%%%%%%%%%%%%%%%%%%%%%%%%%%%%%%%%%%%%%%%%%%%%%%%%%%%%%%%%%%%%%%%%%%%%%%%%%%%%%%%%%%%%%%%%%%%%%%%%%%%%%%%%%%%%%%%%%%%%%%%%%%%%%%%%%%%%%%%%%%%%%%%%%%%%%%%

%%%%%%%%%%%%%%%%%%%%%%%%%%%%%%%%%%%%%%%%%%%%%%%%%%%%%%
%%%%%%%%%%%%%%%%%%%%%%%%%%%%%%%%%%%%%%%%%%%%%%%%%%%%%%

\begin{figure*}
%\hspace{-1cm}
\begin{minipage}[h]{1\linewidth}
\vspace{-0.3cm}
\begin{minipage}[h]{0.5\linewidth}
\begin{center}
\includegraphics[width=0.8\linewidth]{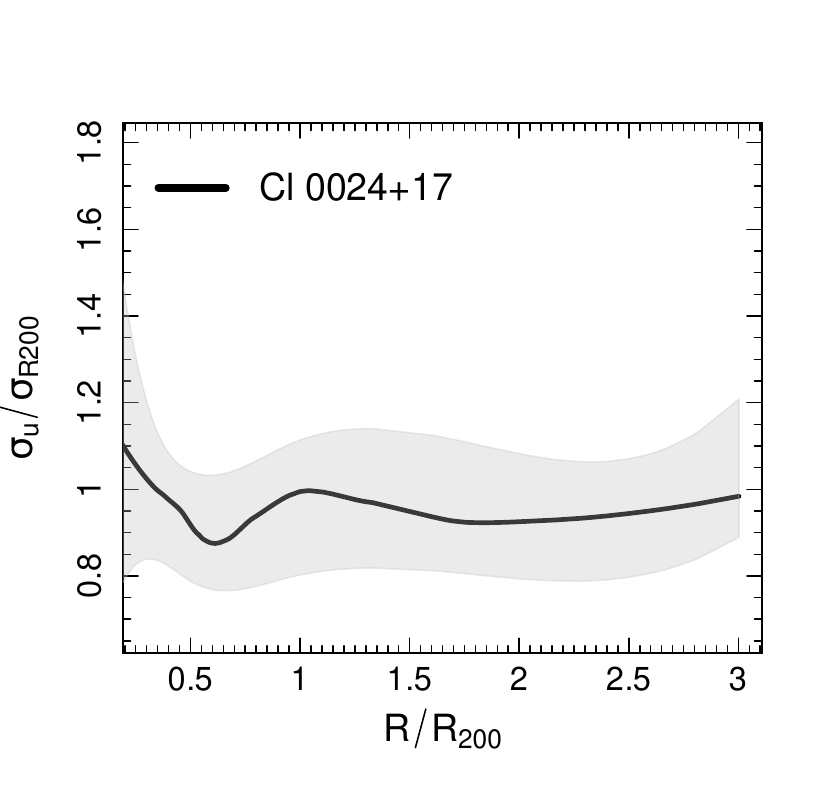} 
%\caption*{}
\label{qwe1}
\end{center} 
\end{minipage}
%\hfill
\vspace{-0.5cm}
\begin{minipage}[h]{0.5\linewidth}
\begin{center}
%\hspace{-3.5cm}
\includegraphics[width=0.8\linewidth]{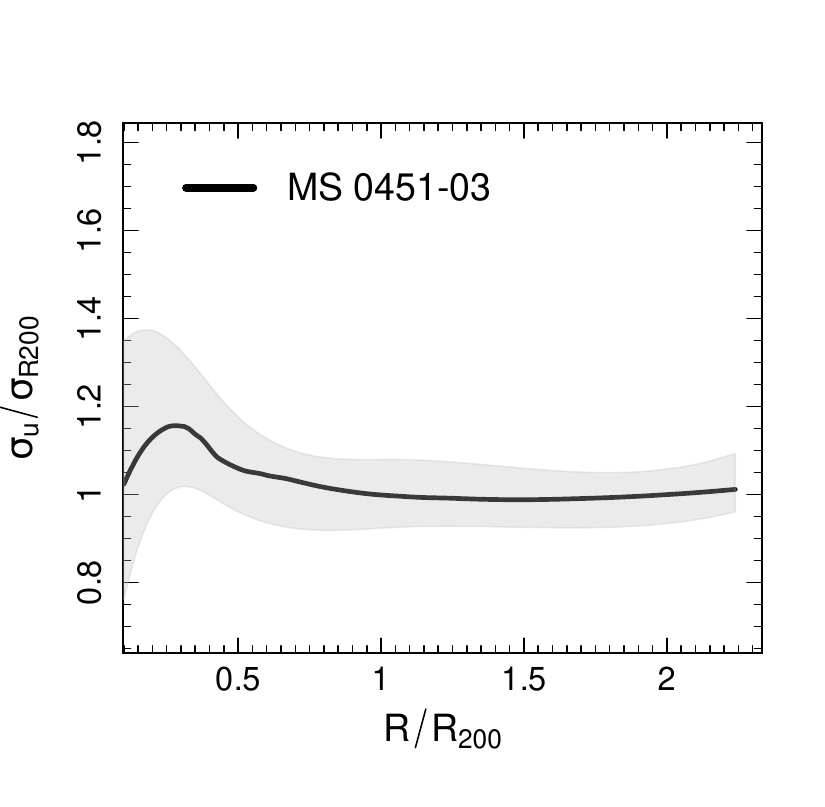} 
%\caption{r1}
\label{qwe1}
\end{center}
\end{minipage}
%\vfill
\vspace{-0.5 cm}
\begin{minipage}[h]{0.5\linewidth}
\begin{center}
\includegraphics[width=0.8\linewidth]{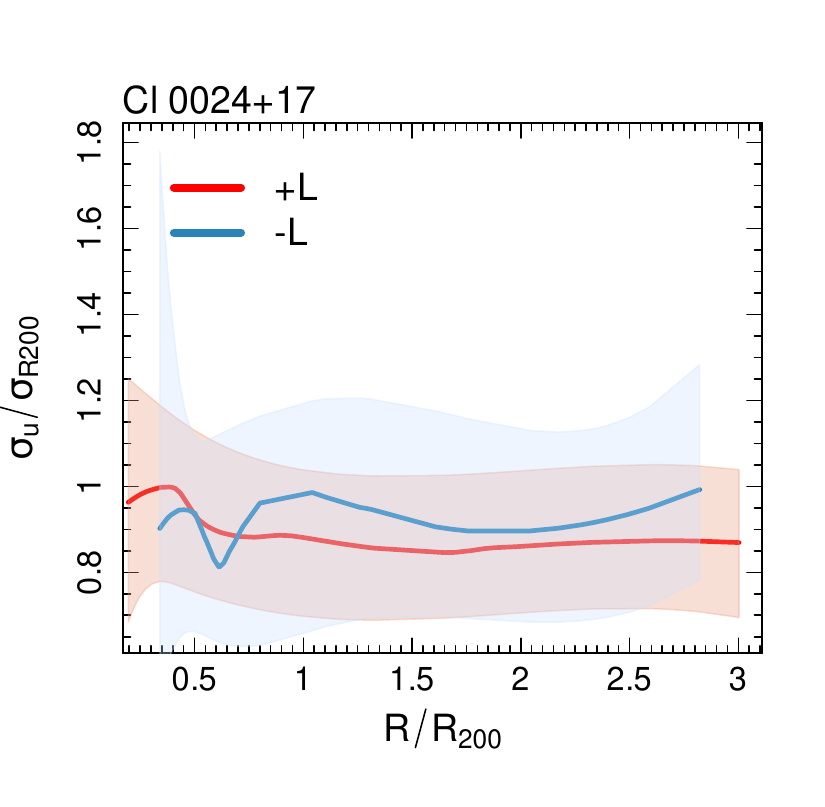} 
%\caption{r1}
\label{qwe1}
\end{center}
\end{minipage}
%\hfill
%\vspace{-0.2cm}
\begin{minipage}[h]{0.5\linewidth}
\begin{center}
\includegraphics[width=0.8\linewidth]{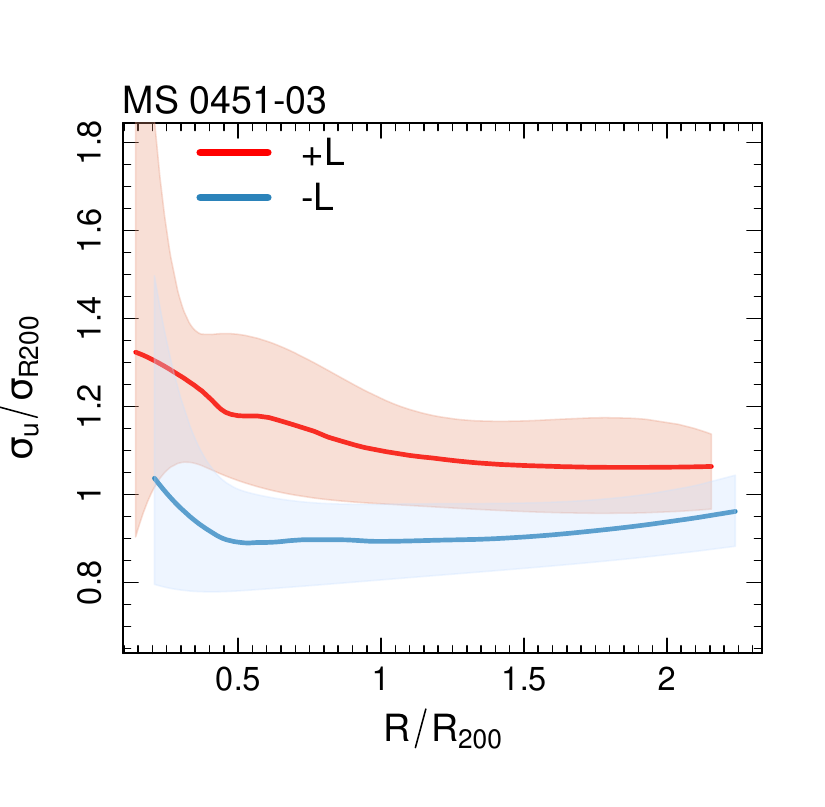} 
%\caption{r1}
\label{qwe1}
\end{center}
\end{minipage}
%\hfill
\begin{minipage}[h]{0.5\linewidth}
\begin{center}
\includegraphics[width=0.8\linewidth]{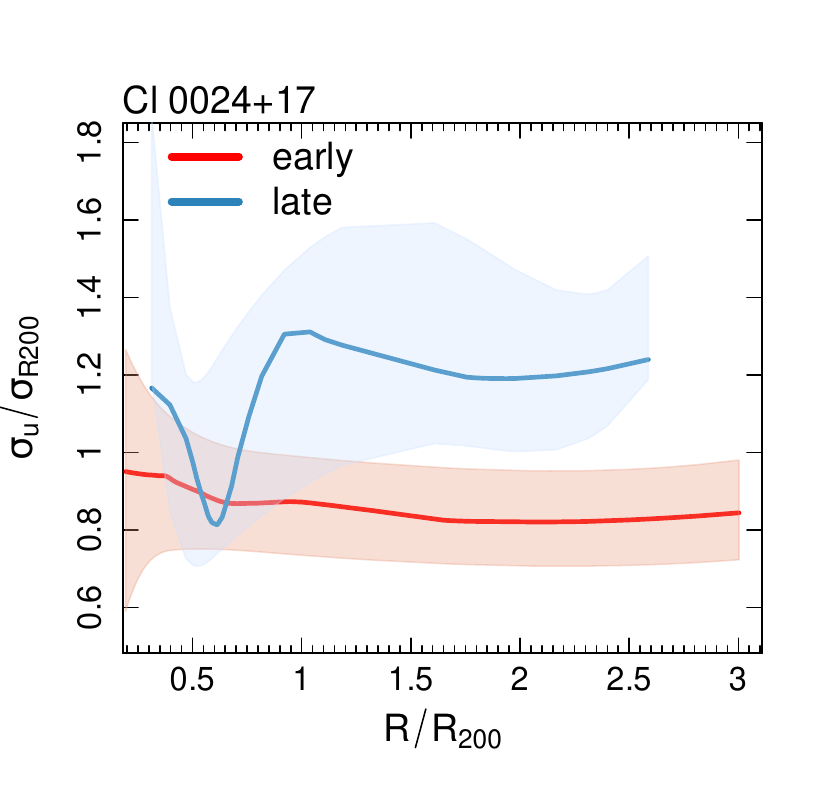} 
%\caption{r1}
\label{qwe1}
\end{center}
\end{minipage}
\hfill
\begin{minipage}[h]{0.5\linewidth}
\begin{center}
\includegraphics[width=0.8\linewidth]{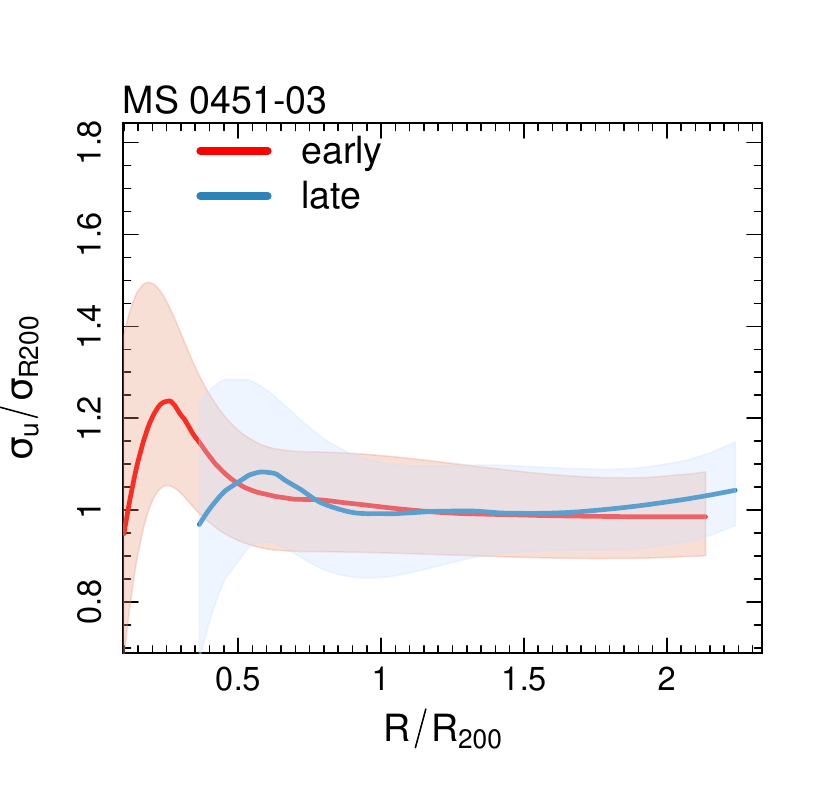} 
%\caption{r1}
\label{qwe1}
\end{center}
\end{minipage}
\vspace{-0.3cm}
\caption{VDPs for Cl 0024+17 (left column) and MS 0451-03 (right column). Overall VDP (solid black line), +L and -L galaxies, in red and blue lines, respectively.
       The early and late components are also represented by respective colors. The VDPs are normalized by the dispersion equivalent to the $R_{200}$ region of Cl 0024+17 and MS 0451-03 defined according to  \protect\cite{1997ApJ...478..462C}.
        }
\label{figvdps}
\end{minipage}
\end{figure*}

In Figure \ref{figvdps} we see the VDPs for Cl 0024+17 (left column) and MS 0451-03 (right column), where the solid black line represents the VDP taking into account all galaxies. The shaded areas around each profile (solid line) are representative of the confidence intervals (90$\%$) obtained through 1000 bootstraps. The red and blue lines below the VDP considering all galaxies in each cluster show the VDPs for the $+L$ and $-L$ and ETGs (early) and LTGs (late) for each cluster. The projected distance is normalized by $R_{200}$. This figure shows that both clusters exhibit totally distinct VDP shapes, not only when using all galaxies but also when examining the different subsamples separated by luminosity or morphology. 

In the case of Cl 0024+17 there is a slight depression in the VDP around $0.5 R_{200}$, followed by an increase, and then an almost flat trend. Usually the presence of depressions in VDPs suggests subgroups experiencing merger, as discussed in several studies, for example, \cite{1996ApJ...472...46M}, \cite{2003ApJ...590..225C}, \cite{2011MNRAS.413L..81R} and \cite{2018MNRAS.473L..31C}. This is reinforced by the presence of at least two substructures close to the virial region, as shown in \cite{moran2007wide} and \cite{2018A&A...612A..17W} which
identified a multiple-image configuration through strong lensing. 
The almost flat trend displayed after the depression in Cl 0024+17 is also in agreement with the \cite{1998ApJ...505...74G} studies schematized in our Figure \ref{fig_schema}, pointing to isotropic velocities. Likewise, \cite{2020A&A...642A.131S} indicates that galaxies with 
isotropic convergent velocity field 
can be associated with an infall pattern.
This results, combined with the non-Gaussianity (\texttt{Mclust} and AD) of the velocity distribution of this system, 
reinforces the understanding that Cl 0024+17 is a system undergoing infall processes.

Also in Cl 0024+17, we observe that the -L component exhibits a slightly higher velocity dispersion compared to the +L component across nearly the entire radial range. This trend is similar to what we find in the VDP of the LTGs sample relative to the ETGs, a characteristic also noted by \cite{2017A&A...606A.108C} for similar morphological components.  Though it's important to acknowledge that the confidence envelopes do overlap, it's also important to make clear that the difference between the averages can still be significant.

The VDPs for the -L and LTGs samples in Cl 0024+17 also show the overall VDP depression, actually more pronounced, while the +L and early components are quite similar to each other and no significant depression is seen.
These findings point to the sense that the -L and LTGs subsamples in Cl 0024+17 are the components responsible for the depression seen in the overall VDP (black solid line) considering all galaxies.

For overall VDP in MS 0451-03, we see a peak in the center, following a slight slope to a predominantly flat shape at radii greater than $1\times R/R_{200}$. This behavior is in accordance with the shape of the VDP shown in the Figure \ref{fig_schema} above, where the initial shape of VDP - considering all galaxies present - in MS 0451-03 indicates a system with isotropic velocities in the central region, that is, with an equal distribution of velocities in all directions in this region. This result becomes more robust when integrated with the analysis of the velocity distribution of MS 0451-03 (through \texttt{Mclust} and AD) whose diagnostic is  Gaussianity for both tests applied. 

The sign of the decreasing shape post-peak, as predicted by \citealt{1998ApJ...505...74G}, points to the presence of galaxies predominantly on radial trajectories.
This conclusion also agrees with the normality tests carried out outside the virial ($R_{200}$) region of MS 0451-03 (see section \ref{veldisp}), since that outside the virial region MS 0451-03 has an NG velocity distribution, probably due to disturbances in the velocity distribution caused by galaxies coming from infall region, bearing in mind that galaxy velocity field
of the infalling regions of clusters is expected to contain a significant radial infall component superposed to other orbits (e.g. \citealt{2016MNRAS.455..127M} and \citealt{2021A&A...655A.103Z}).

For +L and -L subsamples, we see that the velocity dispersion of the galaxies classified as +L being larger than those classified as -L, and with both components following similar behavior. Here, we notice an unexpected inversion in the +L and -L components of the VDPs. A possible explanation for this unusual behavior may be through the mass-luminosity relationship (e.g.~\citealt{lumi1}; ~\citealt{lumi2}; ~\citealt{lumi3}). In view of the fact that more luminous galaxies are often related to higher mass, (according to mass-luminosity relation) and considering that the velocity dispersion is directly influenced by the mass of the systems (e.g. ~\citealt{massa1}; ~\citealt{massa2}; ~\citealt{massa3}), galaxies with higher luminosities (generally, higher masses) may contribute to a higher velocity dispersion.

Other evidence (indirect, in this case) that corroborates this interpretation comes from the comparison of the average values of the [OII] emission lines for both populations. We find that the +L galaxies have an average value of [OII] approximately three times greater than the value displayed for -L galaxies. This may indicate that the +L galaxies has a more significant star formation, probably due to efficiency in accrete more gas at higher redshifts (\citealt{2023A&A...678A..65C}). A possible explanation for this is that the +L galaxies have deeper potential wells  implying higher velocity dispersion values (e.g. ~\citealt{oegerle1990clusters}). %\newpage

Analysing the ETG and LTG components in MS 0451-03, we observe that both exhibit similar shapes as presented by the overall VDP, except that the samples have a small displacement at projected distances from each other, indicating that ETGs exerts greater influence in the initial shape of the overall VDP. 

It is essential to stress out that, although our interpretations regarding the physics implicit in the shape of the VDPs of these two clusters are robust (given the amount of information available in the literature for both clusters), more specific studies on the orbital behavior of galaxies are important for more categorical synthesis.

%%%%%%%%%%%%%%%%%%%%%%%%%%%%%%%%%%%%%%%%%%%%%%%%%%%%%%%%%%%%%%%%%%%%%%%%%%%%%%%%%%%%%%%%%%%%%%%%%%%%%%%%%%%%%%%%%%%%%%%%%%%%%%%%%%%%%%%%%%%%%%%%%%%

\subsection{Projected Phase-Space - PPS}
\label{subpps}

The projected phase-space (PPS) is the representation of the line-of-sight peculiar velocity normalized by the cluster velocity dispersion versus the clustercentric distance normalized by $\rm R_{200}$ - ($| v_{los} | / \sigma_{\rm R_{200}} \times \rm R/R_{200}$). It is a useful visualisation of the internal kinematics of galaxy clusters and may reveal possible effects of spatial and/or kinematic segregation of galaxy populations, e.g., \cite{2010A&A...520A..30M}, 
\cite{2012ApJ...754...97H}, \cite{2013MNRAS.431.2307O}, \cite{2017MNRAS.472..409L},  \cite{2017ApJ...843..128R}, \cite{vm2021} and \cite{andrepps}. After building the PPS for Cl 0024+17 and MS 0451-03 we apply an adaptive kernel technique to create a density map (see, for example, \citealt{venables2013modern}).

%% =======================

\begin{figure}%10.3cm, height=10cm
   \hspace{-0.4cm}
    \includegraphics[height=9.5cm, width=9.5cm]{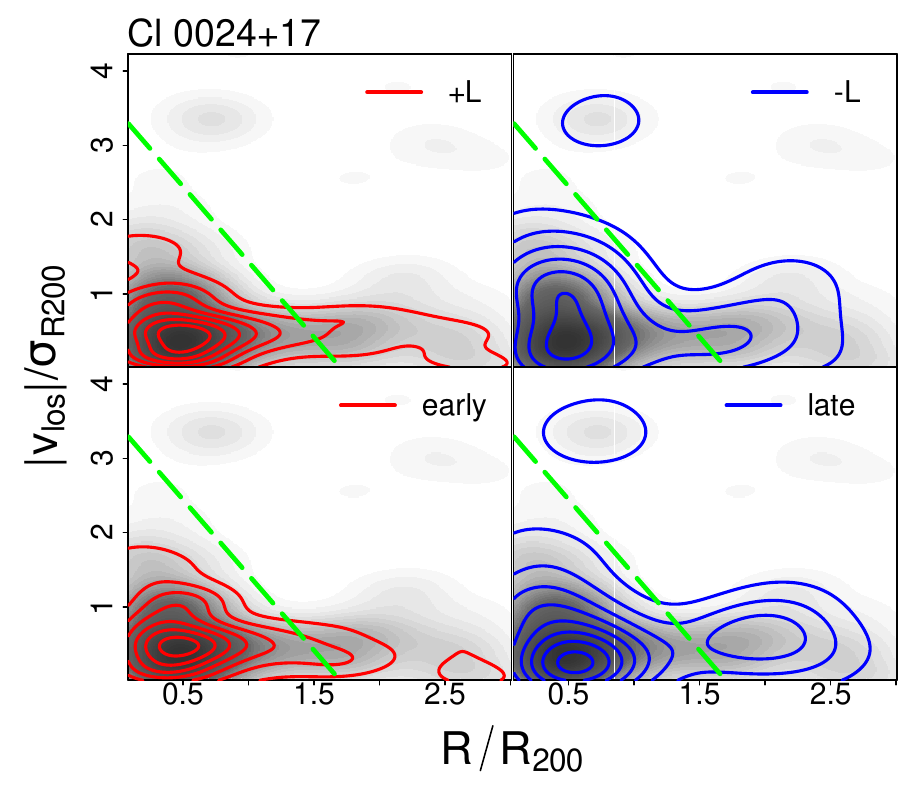} 
    %       \hspace{-0.5cm}
	%\includegraphics[height=9.5cm, width=9.5cm]{PPS_ms.pdf}
%\vspace{-0.2cm}
\caption{
Phase-space for Cl 0024+17.
      In the graphs we see the PPS for Cl 0024+17 where we overlap the isodensity contours for early (and $+L$) (with red contours) and late (and $-L$) (blue contours) populations.
}
\label{pps1}
\end{figure}

%% =======================

\begin{figure}%10.3cm, height=10cm
\hspace{-0.5cm}
        \includegraphics[height=9.5cm, width=9.5cm]{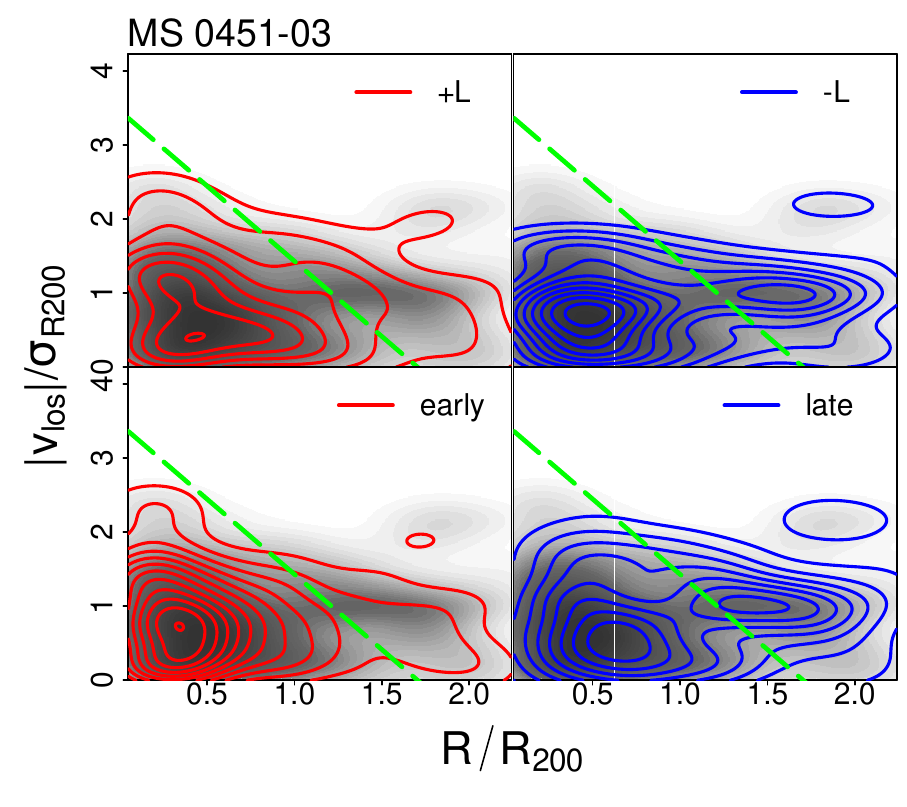}
    %       \hspace{-0.5cm}
	%\includegraphics[height=9.5cm, width=9.5cm]{PPS_ms.pdf}
\caption{Phase-space for MS 0451-03 Similar to the PPS for Cl 0024+17, we overlap the isodensity contours for early (and $+L$) (with red contours) and late (and $-L$) (blue contours) populations.}
\label{pps2}
\end{figure}

%% ========================

In Figure~\ref{pps1} and ~\ref{pps2} we show the PPS of Cl 0024+17 and MS 0451-03 adding to the density maps (in gray-black scale) the isodensity curves for the early, late, +L and -L families (in red and blue colors). In each panel the dashed line in green colour was obtained from the equation $\frac{|V_{\rm los}|}{\sigma}=-\frac{4}{3}\frac{R}{R_{\rm virial}}+2$ (\citealt{2013MNRAS.431.2307O} - hereafter Oman Region). 
Considering projection effects, we made the following assumptions for our observed data, $R_{\rm virial}=\frac{2.5}{2.2}R_{200}$ e $\sigma=\sqrt{3}\sigma_{\rm group}$. The green line in Figures~\ref{pps1} and ~\ref{pps2} can be used as rough discriminator of recent infallers ($\tau <1~Gyr$) and those that are not so recent ($\tau>1~Gyr$) according to \cite{2017MNRAS.467.4410A}.

In the PPS for Cl 0024+17, Figure \ref{pps1}, we note that the isodensity curves for the $+L$ component suggest a higher concentration towards the internal region of the system, unlike the $-L$ population, which is
considerably less concentrated. Also note that this component shows density contours outside the dashed green line, a region that according to \cite{2017ApJ...843..128R} has a significant probability (around 90$\%$) of housing objects called First Infallers. The isodensity contours for the ETGs component are similar to those of the $+L$ population. It also shows a concentration beyond the region delimited by the green line, namely between $1.5R/R_{200}$ and $3.0R/R_{200}$, which may explain the relatively high fraction ($62\%$ and $53\%$, respectively, top Fig. \ref{figesche}) of ETG and +L objects in the outskirts of this system, possibly associated to external sub-clumps identified in this work (see Table \ref{tabela1} section \ref{veldisp}) and also in the study of \cite{moran2007wide}. 

Regarding the LTG and -L populations, especially to LTG, we see that for $|V_{los}|/\sigma_{200} < 2$ the isodensity contours exhibit two peaks that are roughly separated by the green line, while for $|V_{los}|/\sigma_{200} > 2$ we see a concentration of late objects with $R/R_{200} \le 1.5$. 
As stated by \cite{2017ApJ...843..128R}, the high concentration observed inside the dashed green line suggests the presence of galaxies considered Ancient Infallers, objects that have already fallen into the cluster's potential more than 6.5 Gyr ago.
As for the possible concentration with $|V_{los}|/\sigma_{200} >$  2, \cite{2017ApJ...843..128R} finds that there is a 40$\%$ probability of this region being occupied by galaxies recently accrete to the cluster.

For the PPS of MS 0451-03, we see that the +L isodensity curves are more evenly distributed over the entire PPS when compared to those of Cl 0024+17, and the same behavior stands for the ETGs population. The -L component has at least two non-central peaks separated by the green line. The distribution of the isodensity contours for the LTGs population also are similar to those seen in -L group. 

In essence, both PPS of Cl 0024+17 and MS 0451-03 suggest that the clusters present disturbed dynamical configurations. However, it is of paramount importance to explore a bit further how robust are these trends we find by examining the PPS. This is done by using the \texttt{kde.test} function (package \texttt{ks}, \citealt{duong2012closed}) of \texttt{R} (\texttt{R} Development Core Team, http://www.rproject.org) to quantify the statistical significance of the features in the PPS. The algorithm transforms the data points of the PPS into densities through an adaptive kernel and utilize a multivariate nonparametric two-sample test to compare different distributions directly and quantitatively. We test the null hypothesis that the two distributions being compared are taken from the same parent population. 

Table \ref{tabelakde} shows the results. We adopt here, for rejection of the null hypothesis, the confidence level of $90\%$. Our expectation is that in a more evolved system, the ETG (and +L) and LTG (and -L) present significant differences in their phase-space distributions, with ETGs more concentrated into the virialized region (defined by \citealt{2013MNRAS.431.2307O}), and LTGs more spread through the phase-space.  As we can see from  Table \ref{tabelakde}, this initial expectation is confirmed for cluster MS 0451-03 (p-$value = 0.01$), but it is not confirmed for the Cl 0024+17 system (p-$value = 0.36$). In addition, we find that, while the ETGs of both clusters are not significantly different from each other (p-$value = 0.37$), the LTGs distribute very distinctly (p-$value = 0.02$). Note that in the case of Cl 0024+17 the LTGs population seems to have two density peaks beyond the green line (one more centrally located and at high velocities and another more peripheral with low velocities). 

Regarding the segregation in luminosity, i.e., +L and -L, although we visually find differences in the distributions of the PPS, \texttt{kde.test} does not find  significant difference when compared for the same cluster. This is not true when the test is applied to +L and -L populations coming from different systems. We find that the +L subsamples of Cl 0024+17 and MS 0451-03 are statistically differently distributed in the PPS (p-$value = 0.02$). The same is found for the -L populations coming from Cl 0024+17 and MS 0451-03 (p-$value = 0.09$). 

%==============================

\begin{table}
\caption{Results of the comparison between the Cl 0024+17
and MS 0451-03 sub-samples, regarding the distribution in the PPS.}
\begin{tabular}{p{2cm} p{3cm} p{1.5cm}}
\toprule
\textbf{Samples Tested} & \textbf{Clusters} & \textbf{p-value} \\
\midrule
$+L$ \textbf{vs} $-L$ & Cl 0024+17 & $0.26$ \\
$+L$ \textbf{vs} $-L$ & MS 0451-03 & $0.10$ \\

\hdashline
 & Cl 0024+17 \\ 
 $+L$ \textbf{vs} $+L$ & \,\,\,\,\,\,\,\,\,\textbf{vs} & $0.02$ \\ & MS 0451-03 &  \\
\hdashline
 & Cl 0024+17 \\ 
 $-L$ \textbf{vs} $-L$ & \,\,\,\,\,\,\,\,\,\textbf{vs} & $0.09$ \\ & MS 0451-03 &  \\

\midrule

Early \textbf{vs} Late & Cl 0024+17 & $0.36$ \\
Early \textbf{vs} Late & MS 0451-03 & $0.01$ \\
\hdashline
 & Cl 0024+17 \\ 
 Early \textbf{vs} Early & \,\,\,\,\,\,\,\,\,\textbf{vs} & $0.37$ \\ & MS 0451-03 &  \\
\hdashline

& Cl 0024+17 \\ 
 Late \textbf{vs} Late & \,\,\,\,\,\,\,\,\,\textbf{vs} & $0.02$ \\ & MS 0451-03 &  \\
\bottomrule
\end{tabular}
\label{tabelakde}
\end{table}
%%%%%%%%%%%%%%%%%%%%%%%%%%%%%%%%%%%%%%%%%%%%%%%%%
%%%%%%%%%%%%%%%%%%%%%%%%%%%%%%%%%%%%%%%%%%%%%%%%%

% \vspace{-0.5cm}

%%%%%%%%%%%%%%%%%%%%%%%%%%%%%%%%%%%%%%%%%%%%%%%%%
%%%%%%%%%%%%%%%%%%%%%%%%%%%%%%%%%%%%%%%%%%%%%%%%%%%%%%%%%%%%%%%%%%%%%%%%%%%%%%%%%%%%%%%%%%%%%%%%%%

\subsection{Spatial Distribution Analysis}

Previously, we have shown evidence of irregularities in the two clusters analyzed in this work (shape of VDPs and PPS, for example). Considering that due to dynamical effects, irregularities are possible substructures that may be erased on a typical timescale of a few Gyr (e.g. \citealt{melchiorri2006background}), its detection in clusters of galaxies suggests that either we are dealing with still forming systems through large-scale hierarchical processes (e.g. \citealt{1538-4357-451-1-L5}) or we are witnessing a time-dependent assembly process (e.g. \citealt{1538-3881-154-3-96}). Consequently, we should find some irregularities in the spatial distribution of galaxies. Bearing this in mind, we perform a study of the spatial distribution of galaxies in Cl 0024+17 and MS 0451-03.

Considering that due the large random motions and infall velocities of galaxies in the regions around clusters complicate the detection and characterization of substructures through normal group-finding (e.g. \citealt{2024MNRAS.527...23C}), we applied a modified version of the Dressler-Schectman test (\citealt{1988AJ.....95..985D}), introduced by \cite{2017A&A...607A..81B} and tested in detail by \cite{benavides2023dsp}, called DS+. Succinctly, unlike the "standard" version (\citealt{1988AJ.....95..985D} - important technique, but less robust than this), DS+ does not restrict to a fixed number of neighboring galaxies around each member to be scanned.
It considers any possible multiplicity of neighboring galaxies and checks for differences in their kinematics ($\delta_{v}$ and $\delta_{\sigma}$) from that of the cluster as a whole. Then, DS+ assigns a significance to each detected group using Monte Carlo resampling. The identified subgroups that share one or more galaxies with a more significant group are disregarded to avoid overlapping. Groups that are close enough in distance and velocity are merged to avoid fragmentation. Since real groups are not expected to have velocity dispersion larger than clusters, DS+ does not consider as significant velocity dispersion greater than that of the cluster.

DS+ is sensitive to spatially compact subsystems that either have an average
velocity that differs from the cluster mean, or have a velocity dispersion that differs from the global one, or both.
The kinematic parameters used to verify possible deviations in velocity and velocity dispersion of substructures in relation to the host cluster are presented in the following equations:

{\large
\begin{eqnarray}
\delta_{v} &=& N_{g}^{1/2}\, |\,\overline{v_{g}}\,|\,[\,(t_{n}-1)\,\sigma_{v}\,(R_{g})\,]^{-1} 
\label{deltav}
\end{eqnarray}
}

{\large
\begin{eqnarray}
\delta_{\sigma} &=&  \left[ 1 - \frac{\sigma_{g}}{\sigma_{v}(R)}\right]\,\left\{1 - \left[\frac{(N_{g}-1)}{\chi^{+}_{N_{g}-1} } \right]^{1/2} \right\}^{-1}
\label{deltasigma}
\end{eqnarray}
}\\

{\noindent
where $R_{g}$ is the average projected substructure distance from the cluster center, $v_{g}$ is the mean substructure velocity, $\sigma_{v}(R)$ is the
cluster line-of-sight velocity dispersion profile,
and $\sigma_{g}$ is the substructure line-of-sight velocity dispersion. All these measures are taken with galaxy velocities in the cluster rest-frame.} The Student-$t$ and $\chi^{2}$ distributions are used to normalize the
differences in units of the uncertainties in the mean velocity and
velocity dispersion, respectively. 
The $N_{g}$ parameter represents the minimum number of galaxies in a group to be considered as a substructure. Otherwise, $N_{g}$ represents the multiplicity of substructures. For more details, see \cite{benavides2023dsp}.

%%%%%%%%%%%%%%%%%%%%%%%%%%%%%%%%%%%%%%%%%%%%%%%%%%%%%%%%%%%%%%%%%%%%%%%%%%%%%%%%%%%%%%%%%%%%%%%%%%

\begin{figure} %10.3cm, height=10cm
   \centering
   \vspace*{-0.2in}
        \includegraphics[height=8.4cm, width=10cm]{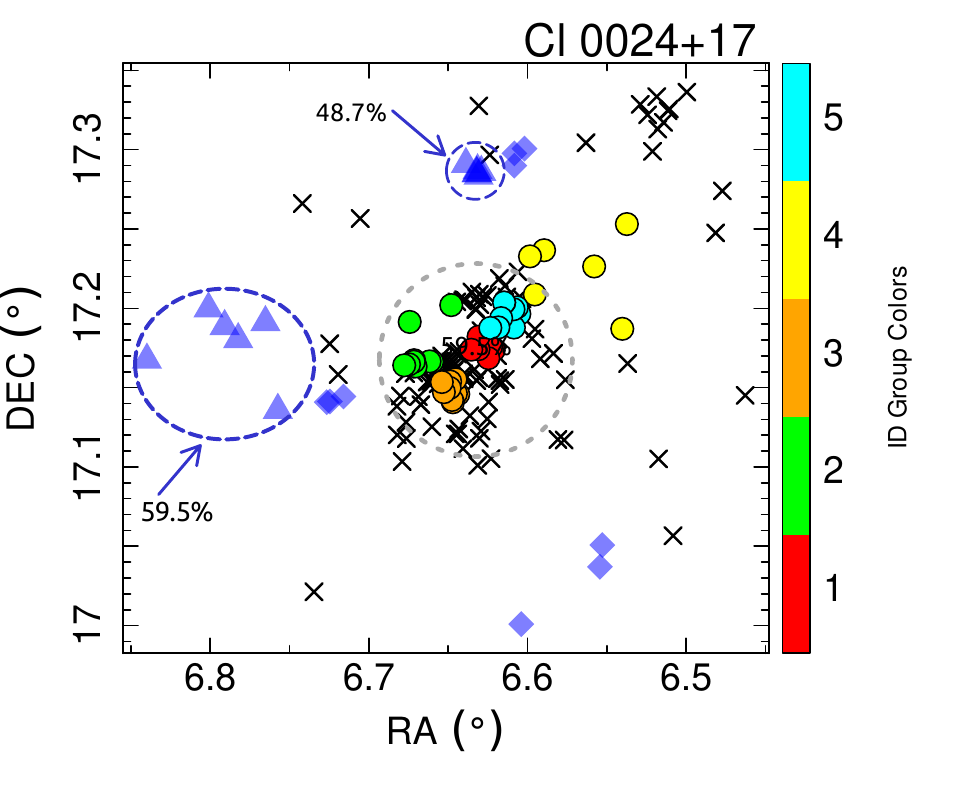}\quad 
    \vspace*{-0.3in}
	 \includegraphics[height=8.4cm, width=9.7cm]{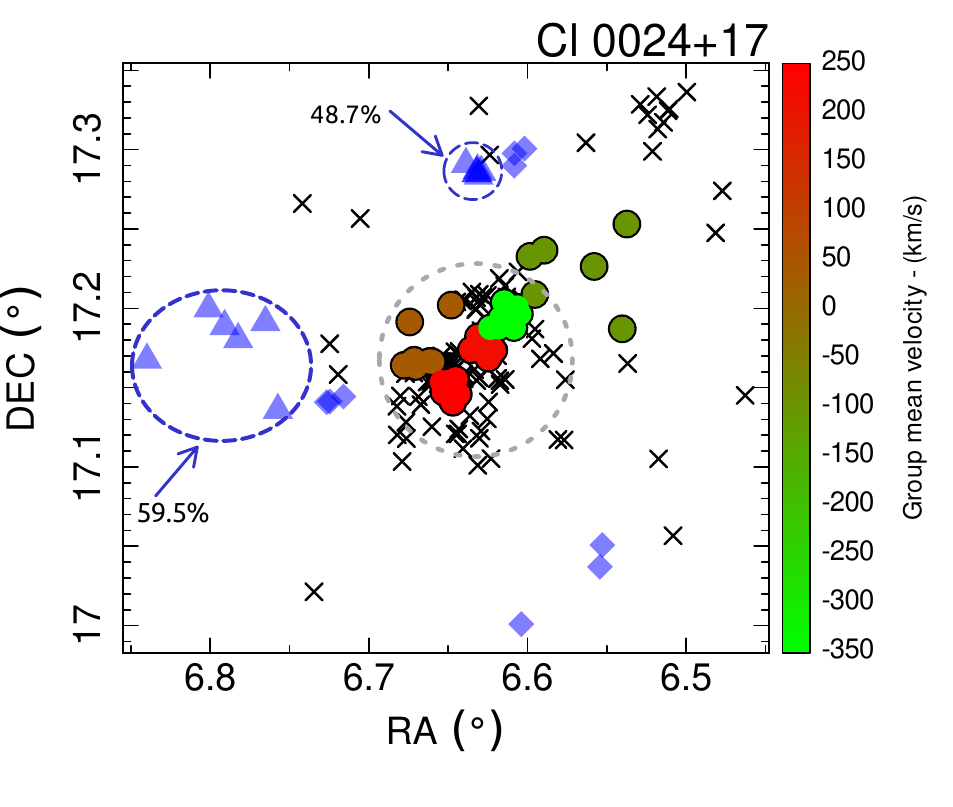}
  % \quad
  % \vspace*{-0.3in}
  % \includegraphics[height=8.4cm, width=8.4cm]{cl0024_pps__mais_subs_ok.pdf}
  \vspace*{0.1in}
       \caption{
RA-DEC distribution of galaxies in Cl 0024+17. 'x' symbols mark field galaxies identified, while colored circles ('o') represent
statistically significant galaxies groups with recovery rate above \protect{$70\%$}, with sizes scaled by \protect{$1-100*p$}, where p is the group's p-value (see Table \ref{cltable}). Blue diamonds denote groups with p-values outside the established statistical reliability. The gray dashed circle indicates the \protect{$R_{200} = 0.95$} Mpc boundary. Blue triangles within dashed circles show substructures with DS+ p-value \protect{$\leq 0.01$} but a recovery rate below \protect{$70\%$}, based on bootstrap significance.}
\label{subppscl}

\end{figure}

%%%%%%%%%%%%%%%%%%%%%%%%%%%%%%%%%%%%%%%%%%%%%

%%%%%%%%%%%%%%% Tabla Cl DS+ %%%%%%%%%%%%

\begin{table*}%
	\centering %
	\caption{Properties of substructures found by DS+ in Cl 0024+17.\label{tab2}}%
	\begin{threeparttable}
		\begin{tabular}{@{\extracolsep{\fill}} C{1.8cm} C{1.5cm} C{1.5cm} C{2.2cm} C{2.5cm} C{2.5cm} C{2.5cm} @{}}
			\toprule
			\textbf{Group ID} & \textbf{N$_{gal}$}  & \textbf{p-value}  & \textbf{Size [kpc]}  & \textbf{Group Color} & \textbf{$\sigma_{grp}$ [km/s]} & \textbf{Bootstrap c.l.} \\
			\midrule
			01 & 06  & 0.001  & 90.44$\pm$16.01  & red & 204.3$\pm$51.02 & 71.2$\%$ \\
			02 & 09  & 0.008  & 101.70$\pm$12.23  & green & 351.12$\pm$24.13 & 75.4$\%$ \\
			03 & 09  & 0.001  & 38.67$\pm$09.11  & orange & 203.78$\pm$32.09 & 78.9$\%$ \\
			04 & 06  & 0.008  & 1033.32$\pm$37.54  & yellow & 217.00$\pm$23.98 & 70.1$\%$ \\
			05 & 09  & 0.005  & 101.30$\pm$23.57  & cyan & 326.37$\pm$23.20 & 86.3$\%$ \\
			\bottomrule
		\end{tabular}
		\begin{tablenotes}
			\item \textbf{Notes}: The 
			\textbf{Group ID} column represents the number of the identified group by DS+, \textbf{N$_{gal}$} the number of galaxies in each substructure, \textbf{p-value} is the smallest p-value between $\delta_{v}$ and $\delta_{\sigma}$ (main condition for a group to be considered as a substructure), \textbf{Size} is the size of the group in kiloparsec,
			\textbf{Group Color} the color of each group with its respective properties,
			$\sigma_{grp}$ the group 
			velocity dispersion calculated using equation \ref{new_vdp} 
			and 
			\textbf{Bootstrap c.l.} is the statistical confidence of each subgroup via bootstrap. The calculated errors are 1$\sigma$. 
		\end{tablenotes}
	\end{threeparttable}
	\label{cltable}
\end{table*}

%%%%%%%%%%%%%%%%%%%%%%%%%%%%%%%%%%%%%%

%%%%%%%%%%% Tabela MS DS+ 

\begin{table*}%
	\centering %
	\caption{Properties of substructures found by DS+ in MS 0451-03.\label{tab2}}%
	\begin{threeparttable}
		\begin{tabular}{@{\extracolsep{\fill}} C{1.8cm} C{1.5cm} C{1.5cm} C{2.2cm} C{2.5cm} C{2.5cm} C{2.5cm} @{}}
			\toprule
			\textbf{Group ID} & \textbf{N$_{gal}$}  & \textbf{p-value}  & \textbf{Size [kpc]}  & \textbf{Group Color} & \textbf{$\sigma_{grp}$ [km/s]} & \textbf{Bootstrap c.l.} \\
			\midrule
			01 & 06  & 0.000  & 74.67$\pm$22.12  & red    & 343.11$\pm$49.03   & $73.7\%$ \\
			02 & 09  & 0.002  & 117.78$\pm$26.17  & green  & 392.12$\pm$42.73   & $79.1\%$ \\
			03 & 12  & 0.001  & 105.93$\pm$16.53  & orange & 436.49$\pm$84.30   & $86.7\%$\\
			04 & 12  & 0.004  & 370.66$\pm$53.44  & yellow & 242.68$\pm$45.14   & $82.8\%$ \\
			05 & 09  & 0.006  & 124.10$\pm$37.36  & cyan   & 519.51$\pm$41.27   & $81.5\%$ \\
			\bottomrule
		\end{tabular}
		\begin{tablenotes}
			\item \textbf{Notes}: The columns have the same interpretations as those described in Table \ref{cltable}.
		\end{tablenotes}
	\end{threeparttable}
	\label{msdstable}
\end{table*}

DS+ offers as a result, for example, the number of galaxies in each group, the p-value of the kinematic parameter considered, the size of the identified substructure, 
average velocity and velocity dispersion of the group.
We estimate the probability of $\delta_{v}$ and $\delta_{\sigma}$ for MS 0451-03 and Cl 0024+17 by comparing them with the corresponding values obtained for 1000 Monte 
Carlo resamplings in which we replace all the cluster galaxy velocities with random Gaussian draws from a distribution of zero mean and dispersion equal to 
$\sigma_{v}(R_{g})$. We considered as substructure only the groups that presented  p-$value$  
(minimum p-$value$ between $\delta_{v}$ and $\delta_{\sigma}$ of the group) $< 0.01$ \citep{benavides2023dsp}. Furthermore, groups must have a number of galaxies 
greater than the minimum multiplicity established in the analysis ($N_{g} = 3$). With these restrictions, we ensure that the substructures are composed only of galaxies 
with statistical significance. On the other hand, the detection of substructures within galaxy clusters is very difficult, considering that the gravitational potential 
of the cluster tends to destroy the infalling groups within a few Gyr.  \cite{benavides2023dsp} has shown that although the completeness of the groups via DS+ is 
relatively low (due to the difficulty of recovering complete large groups), the purity of the substructures is significant ($>60\%$, indicating that some galaxies in 
these substructures are still gravitationally bound).
With this in mind, we assess the statistical reliability of the groups by performing 10000 bootstraps on the velocity distribution, counting how often a predominant 
part of the substructures is recovered. Given the numerical limitations of our sample, we set that only substructures recovered more than $70\%$ of the time are 
considered statistically relevant.

For Cl 0024+17, our results updated its dynamic configuration through the identification of three new substructures in this cluster, reaching the total of five 
substructures.
This is shown in Figure \protect{\ref{subppscl}}, where the upper panel represents the distribution of objects in RA and DEC, with the color bar having a color to 
identify each of the seven groups, while the lower panel represents the same cluster with the color axis scaled according to the mean velocity of each identified groups.
In addition to the two most central substructures 
identified by \cite{moran2007wide} (green and orange groups), we 
found two other groups within $R_{200}$ (cyan and red points), and one more groups out of $R_{200}$ (yellow) possibly falling into the Cl 0024+17 potential well. These 
findings, along with the results from the previous sections, support the scenario of a more dynamically disturbed system. 
This non-relaxed picture, is also described by \cite{2014ApJ...781...24G} when studying the Intra Cluster Light (ICL) of this field. The authors found that the ICL in 
cluster cores of Cl 0024+17 is produced by stars stripped from the halos of their parent galaxies in the cluster potential well - a more frequent event in non-relaxed 
systems, since in these clusters the interactions between individual galaxies can be more chaotic (e.g.~\citealt{coma}, \citealt{2009ApJ...699.1595P}; and references therein).

%%%%%%%%%%%%%%%%%%%%%%%%%%%%%%%%%%%%%%%%%%%%%%%%%%%%%%%%%%%%%%%%%%%%%%%%%%%%%%%%%%%%%%%%%%%%%%%%%%%%%%%%%%%%%%%%%%%%%%%%%%%%%%%%%%%%%%%%%%%%%%%%%%%%%%%%%

\begin{figure} %10.3cm, height=10cm
   %\centering
   %\vspace*{-0.2in}
   \hspace*{-0.8cm}
        \includegraphics[height=8.4cm, width=10cm]{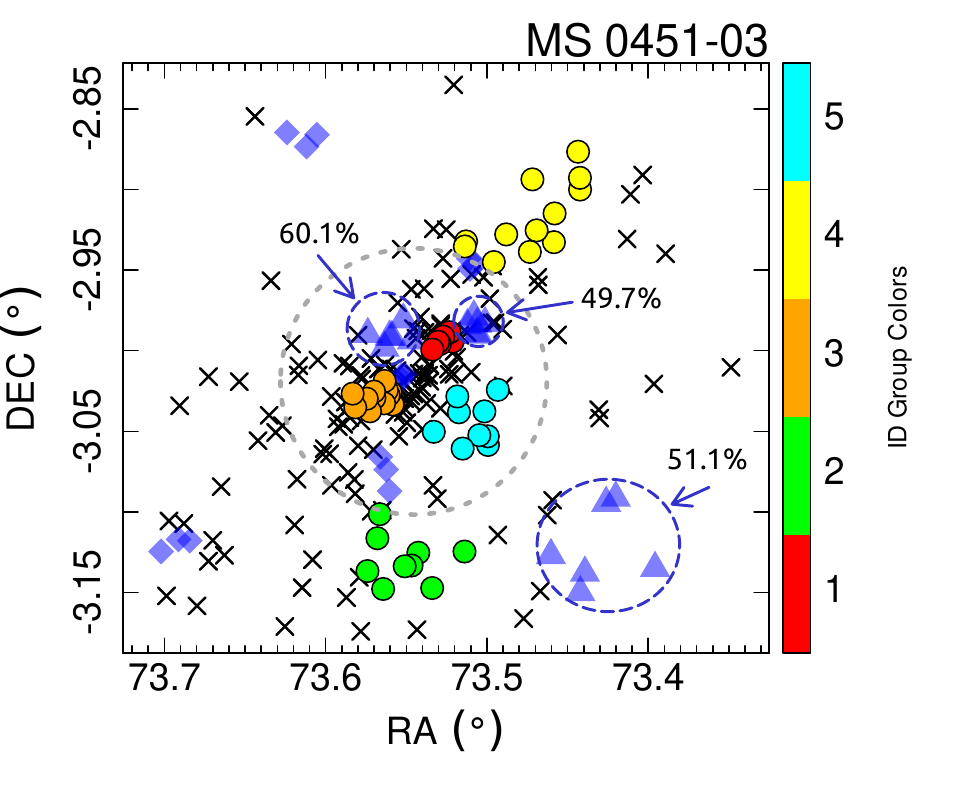}\quad 
    %\vspace*{-0.3in}
    \hspace*{-0.8cm}
	    \includegraphics[height=8.4cm, width=9.7cm]{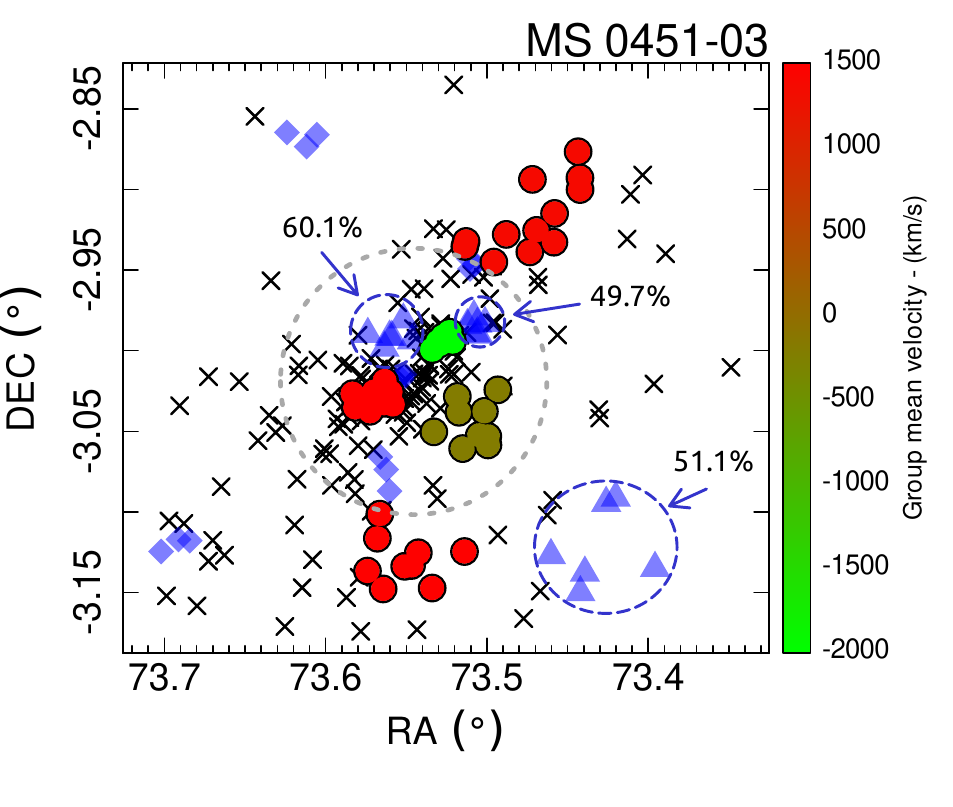}
    
    \caption{RA-DEC distribution for the galaxies of MS 0451-03. The symbols represents the same properties than Cl 0024+17. The size of the colored circles are scaled by the relationship \protect{$1-100*p$}, also similar to what was done in \citet{benavides2023dsp}, where p is the p-value of the group shown in Table \ref{msdstable}. The circle dotted in gray in both plots also has a radius equivalent to \protect{$R_{200} = 1.45$} Mpc.}

\label{fig07}
\end{figure}
%%%%%%%%%%%%%%%%%%%%%%%%%%%%%%%%%%%%%%%%%%%%%%%%%%%%%%%%%%%%%%%%
%%%%%%%%%%%%%%%%%%%%%%%%%%%%%%%%%%%%%%%%%%%%%%%%%%%%%%%%%%%%%%%%

The results for analysis of MS0451-03 are presented in Figure \ref{fig07} and Table \ref{msdstable}. The analysis of substructures also revealed the presence of five 
subgroups with kinematics contrasting with the global state of the cluster. This new result for MS 0451-03 differs from the analysis carried out by \cite{moran2007wide}, 
who did not find any significant subgroup using DS "standard" method. 
The inner region of MS 0451-03 ($<R_{200}$) houses three substructures (see Table \ref{msdstable}), while the outer region of the cluster  exhibits the presence of two 
substructures. These results agree with the combined strong- and weak-gravitational lensing analysis carried out by \cite{2020MNRAS.496.4032T}, which suggests a possible 
bimodal cluster core due to the presence of clumps in the mass distribution of MS 0451-03, identifying it as a post-merger cluster. Furthermore, we also identified groups 
with spurious members in both regions of MS 0451-03,  represented by blue diamonds in the respective plot. 

In both figures, blue triangles inside dashed circles represent subgroups of galaxies that exhibit a significant deviation from the environment, indicated by their 
p-values smaller than 0.01 via DS+. However, when the significance of these groups is verified using the bootstrap, they fail to meet the threshold for satisfactory 
significance. The recovery percentages for these subgroups are labeled around them in the respective figures. It is important to clarify that, while p-values may indicate 
statistical significance by rejecting the null hypothesis, this does not necessarily ensure the stability or reliability of the results, particularly in small sample 
sizes. In such cases, bootstrap analyses provide a more direct assessment of result reliability, offering a clearer understanding of their variability.

\begin{figure}
   \centering
   \vspace*{0.15in}
   % Primeira minipage/figura
   \begin{minipage}{8.5cm}
       \centering
       \includegraphics[height=7.4cm, width=8.5cm]{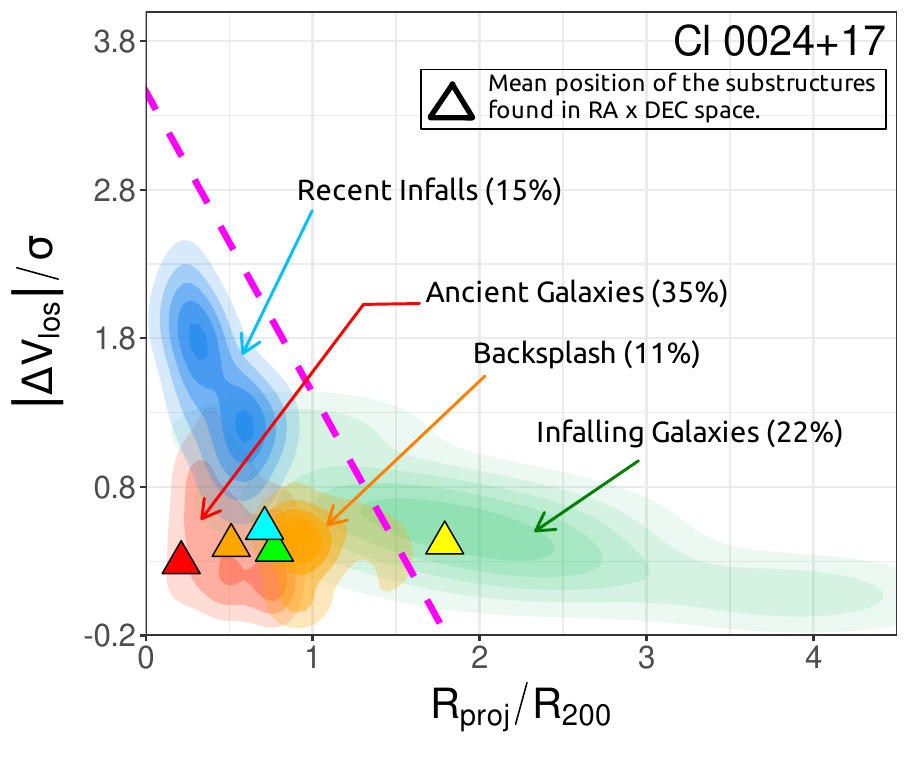}
   \end{minipage}
   % Espaço vertical entre as figuras
   \vspace{1cm} % Ajuste este valor conforme necessário
   
   % Segunda minipage/figura
   \begin{minipage}{8.5cm}
       \centering
       \includegraphics[height=7.4cm, width=8.5cm]{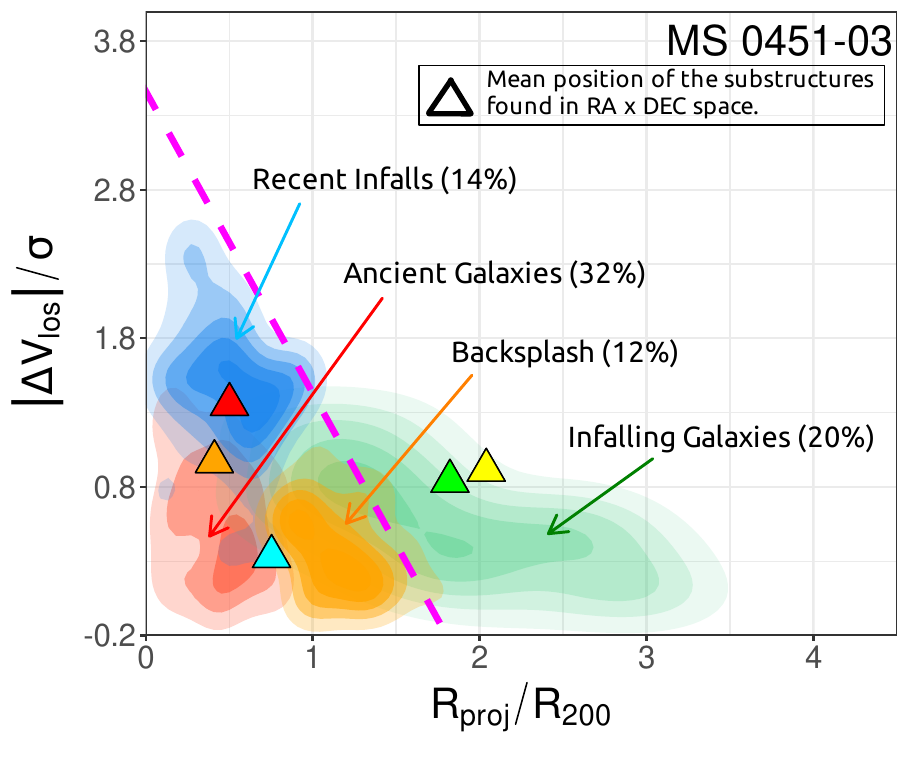}
   \end{minipage}
   \vspace*{0.1in}
   \caption{Projected Phase-Space 
 for Cl 0024+17 and MS 0451-03. In both plots, the shaded regions in red, blue, green and orange represent the regions with the highest probability of existence of 
 galaxies of the type Ancient, Recent Infall, Backsplash and Infalling Galaxies. The percentage of galaxies in each region is also shown. Overlay colored triangles 
 represent the average positions of the substructures found in both clusters. The color of each triangle is in accordance with the colors of the substructures in upper 
 panels of Figures \ref{subppscl} and \ref{fig07}. The dashed pink line is also equivalent to the Oman Region obtained in Figures \ref{pps1} and 
   \ref{pps2}.}
   \label{pps_mais_subs}
\end{figure}

\begin{itemize}
    \item \textit{Substructures and PPS}
\end{itemize}
To understand more about the influence of the substructures found in both clusters, we used \texttt{ROGER} code (\citealt{roger}) to identify in the projected 
phase-space (PPS) (through the reconstruction of the orbits of the galaxies in each cluster) the regions that house different populations of galaxies, namely: 
(i) Cluster members or Ancient Members; (ii) Recent infallers (RIN); (iii) Backsplash galaxies (BS); (iv) Infalling galaxies (IN); and (v) Interlopers (ITL). Briefly, 
the \texttt{ROGER} (Reconstructing Orbits of Galaxies in Extreme Regions) code (\citealt{roger}) is a machine learning technique-based algorithm that classifies galaxies 
based on their position in the projected phase-space (PPS) diagram and their 3D orbits. It uses three different machine learning techniques to classify galaxies in and 
around clusters, according to their projected phase-space position. The code was trained using a galaxy catalogue generated from the MDPL2 cosmological simulation and 
the SAG semi-analytic model of galaxy formation (\citealt{sag}). For each galaxy, \texttt{ROGER} gives as output the probability of being an ancient member, a recent 
infaller, a backsplash galaxy, an infalling galaxy, or an interloper galaxy. This classification is crucial for understanding the impact of different physical mechanisms 
on galaxies and for studying the past trajectories of galaxies in extreme environments such as massive galaxy clusters. The K-Nearest Neighbours method achieved the best 
performance, with a sensitivity of 74$\%$ for classifying cluster galaxies. For more information about the code and definition of the five previously mentioned classes 
of galaxies see \cite{roger}.
After obtaining the PPS using \texttt{ROGER} code, we overlay the substructures found using DS+ with the same \texttt{ID Group Colors} that the substructures identified 
in the upper panels of Figures \ref{subppscl} and \ref{fig07}. We also display the percentage of galaxies contained in each region identified by ROGER.
The result is shown in Figure \ref{pps_mais_subs}.

%%%%%%%%%%%%%%%%%%%%%%%%%%%%%%%%%%%%%%%%%%%%%%%%%%%%%%%%%%%%%%%%%%%%%%%%%%%%%%%%%%%%%%%%%%%%%%%%%%

%%%%%%%%%%%%%%%%%%%%%%%%%%%%%%%%%%%%%%%%%%%%%%%%%%%%%%%%%%%%%%%%%%%%%%%%%%%%%%%%%%

Examining Figure \ref{pps_mais_subs} (top panel), we note the presence of two substructures belonging to the region of galaxies considered ancient members of Cl0024+17. 
Two other substructures occupy a transitional territory between BS galaxies and ancient members, and another the infalling galaxies zone.
As for the PPS of MS0451-03, it is observed that of the five substructures, two are located close to the region that represents the existence of infalling 
galaxies\footnote{
The Infall Galaxies (IN) region is defined by the \texttt{ROGER} code as the region with a significant probability of host young objects, that is, galaxies that do not 
yet have their dynamics completely dominated by the cluster's potential. In other words, galaxies that have never been closer
than $R_{200}$ to the cluster center.}. One substructure  housed in the recent infall region with another moving to ancient galaxies area. Additionally, we note the 
presence of another substructure, located closer to $1R/R_{200}$, situated between the regions of BS galaxies and ancients.
These last two, possibly starting the virialization process with the cluster potential. These results for Cl 0024+17 and MS 0451-03, although pointing to the 
Non-Gaussianity of both clusters, corroborate the proposal that both are at different evolutionary stages. We will discuss these results further in the following section.

\section{Discussion}
\label{sec4}

In this work, we present a detailed dynamical analysis of Cl 0024+17 and MS 0451-03 using a new set of specific tools to access their evolutionary stage.
We consider these two clusters owing to that both constitute a long-term investigation (\citealt{clusters_website11}) aiming to follow the evolution of galaxies in fields 
considered extensive ($\sim$ 10Mpc in diameter). Here, we use systematic indicators of the dynamic stage of clusters that may lead to a more proper interpretation of the 
physical processes in action. Special care was taken in understanding the limits of the catalogs employed (limiting magnitudes and the quality of the morphological information available). 

In this context, while Cl 0024+17 appears to be a more complex system, showing an NG velocity distribution with several gaps in it (Figure \ref{fig06}) - with regions internal and external to R$_{200}$ also exhibing an NG velocity distribution (Table \ref{tabela1}) -, MS 0451-03 displays results compatible with a G velocity distribution, similar to systems that already reached dynamical equilibrium. However, when analyzing the velocity distribution in regions separated by R$_{200}$, it becomes clear that the G diagnosis obtained for the total velocity distribution changes and now the tests indicate G velocity distribution for galaxies internal to R$_{200}$ and NG at radii greater than R$_{200}$. Table \ref{tabela1} presents these findings. In addition, the fractions the brightest objects ($+L$) and the oldest morphological population (ETG) (Figure \ref{figesche}) also endorse MS 0451-03 as more dynamically evolved, with significant dominance of the $+L$ and ETG galaxies for $\rm R \leq R_{200}$. For Cl 0024+17, the region within R$_{200}$ exhibits a high fraction of $+L$ and ETG galaxies, which may be linked to the evolutionary stage of the substructures we identified in the same area. Notice that Cl 0024+17 is supposed to be a colliding system (\citealt{2009ApJ...699.1004Z}). 

At the core of the overall VDP for Cl 0024+17 (black solid line in \ref{figvdps}), we observe a sharp depression inside $R_{200}$, suggesting merger activity. The same behavior is observed by \cite{2018MNRAS.473L..31C} for NG groups. The dynamical interpretation of this VDP behavior is corroborated with theoretical results presented by \cite{1996ApJ...472...46M}. These authors show that in clusters with shallow potential well ($kT < 6.5$ keV)\footnote{Cl 0024+17 has the most recent X-ray temperature measurement of $T \sim 2.6 - 2.8$ keV (\citealt{2009ApJ...699.1004Z}).} the galaxies have a low relative velocity, which enhance merger activity. This scenario, according to \cite{1996ApJ...472...46M}, implies loss or dissipation of orbital energy of the galaxies, leading to central depression in the VDPs. Still, \cite{2018MNRAS.477.5517W} find that recent mergers disturb the gas, reducing the X-ray surface brightness (e.g. \citealt{2011A&A...526A..79E} and \citealt{2017MNRAS.465..213B}). Outside the $R_{200}$ region the VDP shape also takes a behavior equivalent to its temperature distribution, with galaxies behaving as predicted by \cite{1998ApJ...505...74G} (see our Figure \ref{fig_schema}).

In contrast, the overall VDP for MS 0451-03 (a post-merger remnant cluster, \citealt{2020MNRAS.496.4032T}) displays a behavior that is also supported
by X-ray studies, as MS 0451-03 is a system with very high X-ray temperatures ($\sim$ 12.3 keV, \citealt{1998ApJ...502..550D}). Its VDP shape finds endorsement not only in
\cite{1998ApJ...505...74G} (Fig. \ref{fig_schema}), but also in \cite{1996ApJ...472...46M}. This latter, finds that in deep potential well systems ($kT\geq8$ keV), the high relative velocity of galaxies do not allow aggregations. In such cases, the initial VDP shape should be predominantly decreasing. These findings support a non-relaxed scenario for MS 0451-03 (although with signs of being centrally less disturbed than Cl 0024+17), i.e., considering that higher masses and temperatures are more common in NG systems (e.g. \citealt{1538-3881-154-3-96}). Thus, given that in NG systems dynamical perturbations can occur predominantly by merger activity or significantly high infall rate of galaxies in the periphery of the clusters (e.g. \citealt{1538-3881-154-3-96}), we conclude that this latter possibility is most likely to occur in MS 0451-03, since the overall VDP shape do not indicate possible merger activities in the central region of the cluster. 

As far as the PPS analysis is concerned (Figures \ref{pps1} and \ref{pps2}), we find that ETG (and +L) and LTG (and -L) galaxies present significant differences for MS 0451-03 (Table \ref{tabelakde}), contrary to that observed for CL 0024+17, in which the components in morphology and luminosity do not exhibit significant difference in its distribution throughout the PPS.
Complementary to these findings, the analysis of substructures carried out using DS+ (with validation via bootstrap) reflects the complexity involved in the study of clusters at high redshifts. We show that Cl 0024+17 is a cluster with four substructures in its central region, instead of just two as proposed, by \cite{moran2007wide}. In the regions external to $R_{200}$ (extending to the limit of $\sim$ $4 \times R_{200}$) we find one subgroup with high statistical significance, not previously identified. Now, discussing the developments that emerge from this dynamic proxy for MS 0451-03, it is observed that this cluster has a number of substructures equal to CL 0024+17, that is, demonstrating similar uniform spatial distribution. While the central region in MS 0451-03 exhibits the presence of three substantially massive substructures (considering the values of velocity dispersion of these groups in the cluster reference - Table \ref{msdstable}) the region external to $R_{200}$ (with a limit of up to $\sim$ $3.5 \times R_{200}$) suffers the influence of two subgroups probably not yet fully incorporated to the main cluster. 

We emphasize that these findings for MS 0451-03 are unheard when only the kinematics of the galaxies are considered, since the analysis performed by \cite{2020MNRAS.496.4032T} combines gravitational lensing techniques with stellar mass analysis to estimate identified regions of overdensity as substructures. One point to understand is how to reconcile the fact that MS 0451-03 has a significant presence of substructures and still exhibits a G velocity distribution. A possible explanation incorporating both dynamic indicators arise when considering that this central area, hosting approximately $\sim70$$\%$ of the observations, has a VDP indicating that the region is dominated by galaxies with isotropic velocities (see Section 3.2). As a consequence, in a large number of observations, the random fluctuations in velocity directions (attributable to the isotropic nature of the environment) tend to equalize, resulting in a velocity distribution that approximates Gaussian, according to the central limit theorem \citep[e.g.][]{Smith1985RelativeVO}.

Furthermore, we verify the average distribution of these substructures in specific regions of the PPS, derived from machine learning analysis. Through these results (Figure \ref{pps_mais_subs}), we infer that while Cl 0024+17 does not have any substructure in the Recent Infalls region, MS 0451-03 contains at least two groups in the same corresponding region. In this case, according to the definition of Recent Infalls\footnote{Galaxies that have crossed $R_{200}$ only once on their way in the past 2 Gyr} as in \cite{roger}, the substructures at this location (having already crossed the $R_{200}$ region) may be in the process of disruption of their gravitational potential by the central potential of the cluster, implying a future evolution towards ancient-type galaxies, which have already relaxed in the gravitational potential of the cluster (red region - where only the presence of one substructure is noted (transition zone of BS galaxies) in MS 0451-03 and four are verified in Cl 0024+17 (with two overlapping with the  orange zone, suggesting possible BS contamination in these subgroups), in agreement to its NG velocity distribution for this region). Another significant result from this additional study indicates that both clusters have their outer regions
influenced by at least one Infalling subgroups (galaxies that have never been closer than $R_{200}$ to the cluster centre). These results suggest that both clusters experience an off-balance moment of their assembly history process. 

\begin{figure}
   \centering
   \vspace*{0.15in}
   % Primeira minipage/figura
   \begin{minipage}{8.5cm}
       \centering
       \includegraphics[height=8cm, width=10cm]{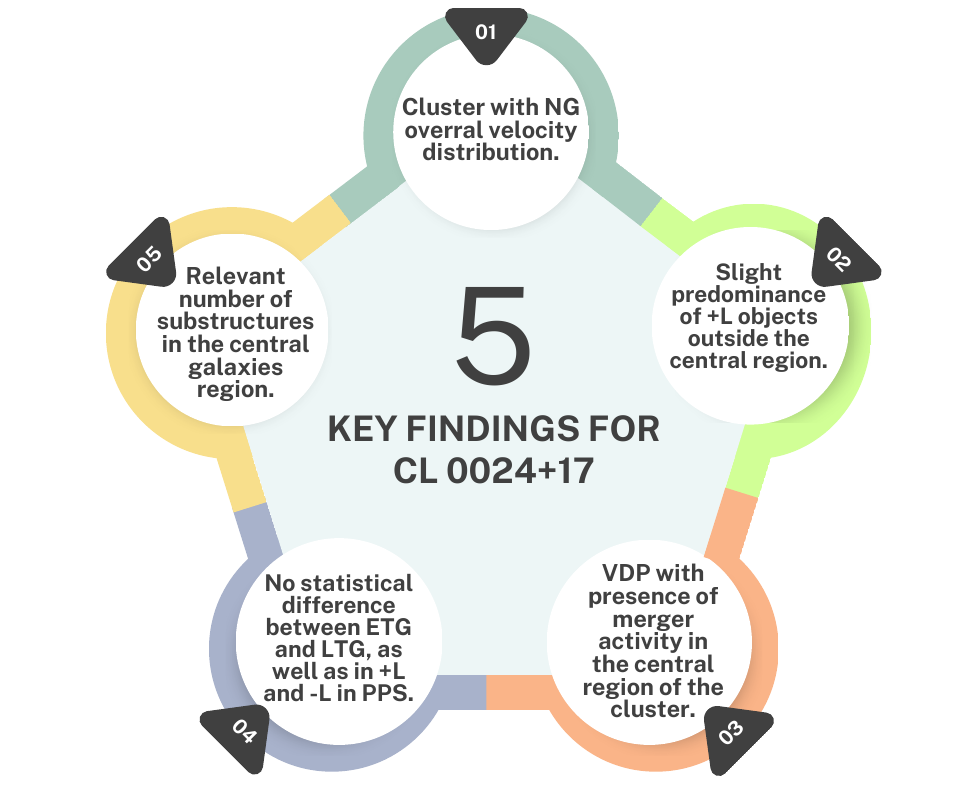}
   \end{minipage}
   % Espaço vertical entre as figuras
   \vspace{1cm} % Ajuste este valor conforme necessário
   
   % Segunda minipage/figura
   \begin{minipage}{8.5cm}
       \centering
       \includegraphics[height=8cm, width=10cm]{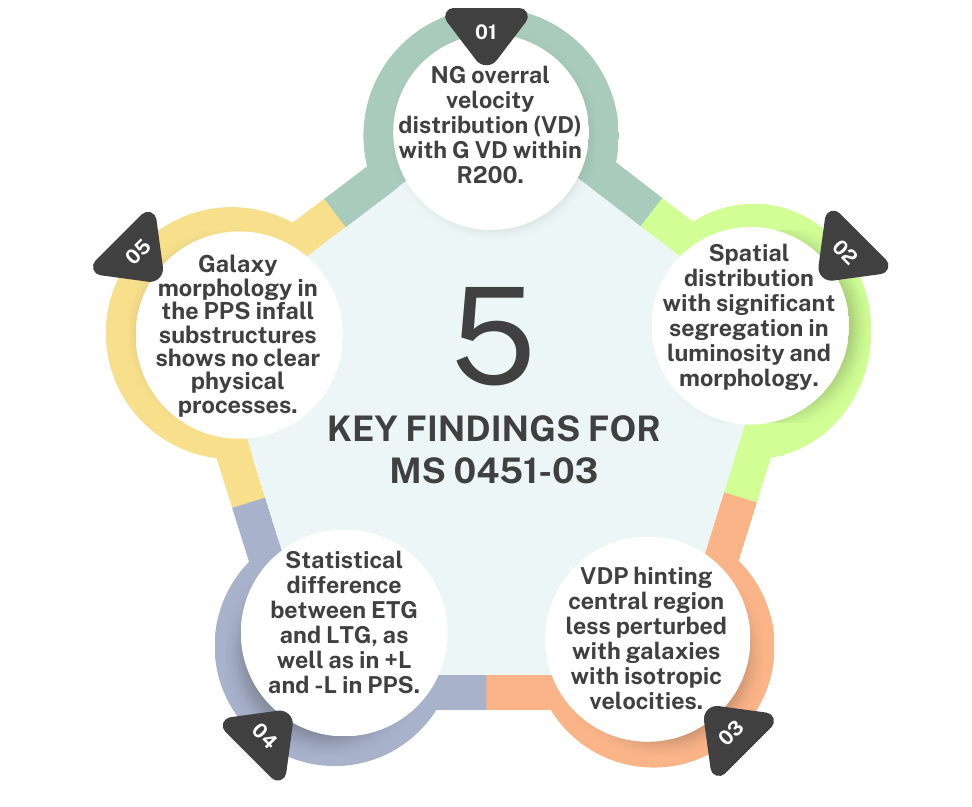}
   \end{minipage}
   \vspace*{0.1in}
   \caption{Summary of some of the characteristics findings in the two clusters analyzed.}
   \label{summary}
\end{figure}

At this stage, driven by observations of \textit{pre-processing}  occurring within galaxy subgroups (e.g. \citealt{2004PASJ...56...29F}, \citealt{2013MNRAS.431L.117M}, 
\citealt{2018MNRAS.479.2328O} at $z\sim0.4$ and   \citealt{2024MNRAS.527L..19L} out to the turnaround radius of clusters) and given theoretical models indicating that 
massive clusters accreted $\sim 40\%$ of their galaxies from infalling groups, generally located at large clustercentric radii ($2 < R/R_{200} < 3 $, e.g., 
\citealt{2013ApJ...770...62D}, \citealt{2014MNRAS.442..406H}, \citealt{2015ApJ...806..101H} and \citealt{2020MNRAS.498.3852B}), we estimate the distribution of the 
morphological population of groups in the region of Infalling Galaxies in both PPS. Based on our findings, it is noted that within the infalling zone of MS0451-03, 
$\sim 40\%$ of the galaxies within the substructures are classified as ETG, while $60\%$ are categorized as LTG. In contrast, Cl 0024+17 exhibits markedly different 
proportions, with the majority ($\sim 83\%$) of their objects falling into the ETG category. With due numerical reservations, this predominance of ETG within the 
substructures located in the infalling region of Cl 0024+17 may be linked with the environment defined by this subgroup that probably quenched most of these objects, 
transforming them into passive systems. As highlighted by \cite{2016MNRAS.461.1202J} and also by \citealt{Sarron2019Pre-processing}, interactions among galaxies are more 
common in small-scale groups, often resulting in a subsequent quenching scenario. Thus, since passive galaxies, which are characterized by their lack of ongoing star 
formation, are frequently ETG (e.g. \citealt{Gobat2012THEEE} and \citealt{2022MNRAS.509..567S}) we can suggest, with restrictions due to the numerical limitation of the 
substructures, that this high fraction of ETGs can be justified by the action of possible pre-processing mechanisms before these galaxies enter the cluster environment. 
Finally, it is of paramount importance to stress the fact that only when more detailed studies like this are conducted on larger samples we will have a better 
understanding of the dynamics of clusters. The diagram depicted in Figure \ref{summary} concisely illustrates several key findings.

%%%%%%%%%%%%%%%%%%%%%%%%%%%%%%%%%%%%%%%%%%%%%%%%%%%%%%%%%%%%%%%%%%%%%%%%%%%%%%%%%%%%%%%%%%%%%%%%%%

\newpage



%\bibliographystyle{mnras}
\section*{Acknowledgements}
We thank the anonymous referee for the useful comments that
greatly improved this paper.
APC thanks Berguem Paula S.A., CEEP Nelson Schaun and Thiago Marcel .S.S. for the support.
ALBR thanks the support of CNPq, grant 316317/2021-7 and FAPESB INFRA PIE 0013/2016.

\section*{DATA AVAILABILITY}
The galaxy clusters used in this work are public available. However, we are willing to provide - upon reasonable request - the separate lists we have created.

\bibliographystyle{mnras}
\bibliography{mnras_edited_final_version_re_sub.bib} % if your bibtex file is called example.bib

% % Alternatively you could enter them by hand, like this:
% % This method is tedious and prone to error if you have lots of references
% \begin{thebibliography}{99}
% \bibitem[\protect\citeauthoryear{Author}{2012}]{Author2012}
% Author A.~N., 2013, Journal of Improbable Astronomy, 1, 1
% \bibitem[\protect\citeauthoryear{Others}{2013}]{Others2013}
% Others S., 2012, Journal of Interesting Stuff, 17, 198
% \end{thebibliography}

%%%%%%%%%%%%%%%%%%%%%%%%%%%%%%%%%%%%%%%%%%%%%%%%%%

%%%%%%%%%%%%%%%%% APPENDICES %%%%%%%%%%%%%%%%%%%%%
% 
% \appendix
% 
% \section{Some extra material}
% 
% If you want to present additional material which would interrupt the flow of the main paper,
% it can be placed in an Appendix which appears after the list of references.

%%%%%%%%%%%%%%%%%%%%%%%%%%%%%%%%%%%%%%%%%%%%%%%%%%

% Don't change these lines
\bsp	% typesetting comment
\label{lastpage}
\end{document}